\documentclass[12pt]{article}
\begin{document}
\input epsf
\def\be{\begin{equation}}
\def\bea{\begin{eqnarray}}
\def\ee{\end{equation}}
\def\eea{\end{eqnarray}}
\def\d{\partial}
\def\eps{\epsilon}

\begin{flushright}
OHSTPY-HEP-T-01-005\\
hep-th/0103169
\end{flushright}
\vspace{20mm}
\begin{center}
{\LARGE Three-point functions for $M^N/S^N$
orbifolds with
${\cal N}=4$ supersymmetry}
\\
\vspace{20mm}
{\bf  Oleg Lunin  and  Samir D. Mathur \\}
\vspace{4mm}
Department of Physics,\\ The Ohio State University,\\ Columbus, OH 43210, USA\\
\vspace{4mm}
\end{center}
\vspace{10mm}
\begin{abstract}

The D1-D5 system is believed to have an `orbifold point' in its
moduli space where its low energy theory is a
${\cal N}=4$ supersymmetric sigma model with target space
$M^N/S^N$, where $M$ is  $T^4$ or $K3$. We study
correlation functions of chiral  operators in  CFTs arising from such a theory.
We construct a basic class of  chiral operators from twist fields of 
the symmetric group and the
generators of the superconformal algebra. We find
explicitly the 3-point functions  for these chiral fields at large 
$N$; these expressions are
`universal' in that they are  independent of the
choice of $M$.  We observe that the result
is a significantly simpler expression  than the corresponding
expression for the bosonic theory based on the same orbifold target space.

\end{abstract}
\newpage

\section{Introduction}
\renewcommand{\theequation}{1.\arabic{equation}}
\setcounter{equation}{0}

The D1-D5 system has been of great interest in recent developments in
string theory.
The system is described by a collection of $n_5$  D5 branes which are
wrapped on a
4-manifold $M$ (which can be a $T^4$ or a $K^3$) and $n_1$ D1 branes
parallel to the
noncompact direction of the D5 branes and bound to them.  This system  has been
very  important for issues related to black holes, since it yields,
upon addition of
momentum excitations,  a supersymmetric configuration which has a
classical  (i.e. not
Planck size) horizon.  In particular, the Bekenstein entropy computed
from the classical
horizon area agrees with the count of microstates for the extremal and near
extremal black holes \cite{stromvafa}. Further,   the low energy  Hawking
radiation from the  hole can be understood in terms of a unitary
microscopic process, not only qualitatively but also
quantitatively, since one finds an agreement of spin dependence and
radiation rates between the semiclassically computed
radiation and the microscopic calculation \cite{dasmathur}. The $AdS/CFT$
correspondence conjecture  gives a duality
         between string theory on   a spacetime  and a certain
conformal field theory (CFT) on the
boundary of this spacetime \cite{mal}. The D1-D5 system gives a CFT
which is dual to the
spacetime
$AdS_3\times S^3\times M$.

While it
is possible to
use simple models for the low energy dynamics of the
D1-D5 system when one is computing the coupling to massless modes of
the supergravity theory (as was done for example in the computations of Hawking
radiation from the D1-D5 microstate), it is believed  that the exact
description of this CFT
must be in terms of  a sigma model with target space being a
deformation of the orbifold
$M^N/S^N$, which is the symmetric orbifold of $N$ copies of
$M$.  (Here $N=n_1n_5$,  and we must take
the low energy limit of the sigma model to obtain
the desired CFT.) In particular we may consider the `orbifold point'
where the target space is exactly the orbifold
$M^N/S^N$ with no deformation. It was suggested in
\cite{sw}\cite{others} that this CFT does correspond to a certain point in
the moduli space of the D1-D5 system.

Thus this orbifold theory would be dual to string
theory on
$AdS_3\times S^3\times M^4$, but at this orbifold point the string theory is
expected to be in a
strongly coupled domain where it cannot be approximated by
tree level  supergravity on a smooth background.   Recall that the
Yang-Mills theory
arising from D3 branes is dual to string theory on $AdS_5\times S^5$.
The orbifold
point of the D1-D5 system can be considered   the analogue of free N=4
supersymmetric Yang-Mills theory, since it is the closest we get to a
simple theory on
the CFT side. Interestingly, it was found \cite{threepoint} that
three point functions at
large N in weakly coupled 4-d Yang-Mills theory arising from D3
branes were equal to
the three point functions arising from the dual supergravity theory,
even though the
supergravity limit of string theory corresponded to {\it strongly}
coupled Yang-Mills. It
would be interesting to ask if there is any similar `protection' of
three point function in
the  D1-D5 system.

In this paper we find three point functions of   a basic class of 
chiral operators in
the orbifold theory.
       The orbifold group that we have  is $S^N$, the permutation
group of
$N$ elements. This group is nonabelian, in contrast to the
cyclic group $Z_N$ which has been studied more extensively in the
past for computation of correlation functions in orbifold
theories \cite{dixon}. Though there are some results in the literature for
general orbifolds \cite{vafaorb}, the study of nonabelian orbifolds
is much less
developed than for abelian orbifolds. It turns out however that the
case of the $S^N$ orbifolds has its own set of
simplifications which make it possible to develop a technique for
computation of correlation functions for these theories. In \cite{our}
a method was developed to compute the
correlation functions for twist operators in the bosonic CFT that
emerges from sigma models with such orbifolds.  The
essential point in that computation  was that for the permutation
group the following simplification emerges. As in
any orbifold theory, one can `undo' the effect of the twist operators
by passing from the space where the theory is
defined  to a covering Riemann surface where fields are single
valued.  For orbifolds of the group $S^N$,   the path
integral with twist insertions becomes an {\it unconstrained} path
integral with no twist insertions (for one copy of
the manifold
$M$) on this covering surface.
         (Such a simplification would not happen if we pass to the covering
surface for a general orbifold group; one would
remove the twists by  going to the cover but the values of the fields
in the path integral at one point on this
cover could be constrained to be related to their values at other
points on the cover.) The path integral on the covering space was
then computed by
using the conformal anomaly of the CFT.

In this paper we extend the calculations of \cite{our} to the case of
theories with ${\cal N}=4$ supersymmetry. Thus we
would obtain in particular results for the  above mentioned `orbifold
point' of the D1-D5 system.  In order to be able
to apply the formalism for any $T^4$ or $K3$ manifold that can appear
in the description of the  D1-D5 system, we
construct a basic class of  chiral operators in the theory in an abstract way,
using only the definition of a twist operators of the
permutation group, and the form of the superconformal algebra
of the $N=4$ CFT. We then compute explicitly
the 3-point function of chiral operators,  in the limit of large $N$,
where the surviving contribution comes from the
case where the covering surface is a sphere.  We observe that the
result is significantly simpler than the
corresponding result in the bosonic theory.

There are several earlier works that relate to the problem we are studying, in
particular \cite{vafaone, frolov, bantay, jevicki, mihailescu, halpern,
Argurio}.

The plan of this paper is as follows. In section 2 we construct the
chiral operators corresponding to twist insertions. In section 3 we
       review briefly the method of \cite{our} for computing correlators
of twist insertions, and explain its extension to the supersymmetric case.
Section 4 discusses correlation functions of currents
that are needed to extend the bosonic computation to the
supersymmetric case, and also computes the 2-point functions.  In
section 5 we compute a simple case of the 3-point functions of these
chiral operators,
and in section 6 we do the general case.  Section 7 is a
discussion.

\section{Constructing the chiral operators $\sigma_n^\pm$}
\renewcommand{\theequation}{2.\arabic{equation}}
\setcounter{equation}{0}

\subsection{Twist operators in the bosonic theory}

Let us start with the bosonic theory, and look at the
definition of twist operators of the theory. We follow the notation
used in  \cite{our},
where the construction is discussed in more detail.   The CFT is
defined over the $z$
plane.  Over each point
$z$ the configuration in the target space is specified by an N-tuple of
coordinates
$(X_1, \dots X_N)$, where $X_i$ is a collective symbol for the
coordinates of a point in the $i$th copy of the manifold
$M$.  The fact that the target space is the orbifold of $M^N$ by the
symmetric group $S^N$ means that  the point
$(X_1, \dots X_N)$  of $M^N$ is to be identified with the points
obtained by any permutation of the
         $X_i$.  First consider the path integral defining the theory in the
absence of any twist operators.  We can integrate
over the $X_i$  independently, without imposing the above
identifications, to obtain a partition function that will be
$N!$ times the partition function $Z$ that would be obtained if we
did take into account the identifications
\be
         \prod_i \int D[X_i] e^{-S(X_i)}\equiv Z_0^N  = N! Z
\label{none}
\ee
where $Z_0$ is the partition function when the target space is just
one copy of $M$.
In the above relation we have assumed that the contribution of the
points where two or
more of the coordinate sets
$X_i$ become equal is of measure zero in the path integral. (Thus note
that (\ref{none}) would not be true if the manifold
$M$ was replaced by a target space that had a finite number of
points; the ignored configurations would then not
have measure zero.\footnote{ Alternatively we can adopt (\ref{none}) as part
of the definition of the theory, taking two copies of the point
$(X_1, \dots X_N)$ if $X_1=X_2$ etc., so that (\ref{none}) is
true by construction. When the target space is a manifold the
configurations affected are of measure zero. })

To define twist operators consider an element of the permutation
group in the form of a `single cycle'
$(1,2,\dots n)$.  (Operators based on a product of two cycles can be
regarded as two single cycle operators placed at
the same point.) To insert a twist operator for this element of the
permutation group at the point
$z=0$ we cut a hole
$|z|<\epsilon$ around this point. As we go around this hole
counterclockwise, we let the first copy of $M$ go over into
the second copy of $M$, and so on, returning to the first copy after
$n$ revolutions. Thus the twist operator changes
the boundary conditions around $z=0$.  The correlation function of
twist operators $\sigma_i$ located at points
$z_i$ is then defined to be the ratio of the path integral performed
with these twisted boundary conditions to the
path integral (\ref{none}):
\be
\langle \sigma_1^\epsilon(z_1)\dots \sigma_k^\epsilon(z_k)\rangle \equiv
\frac{\int_{\rm twisted} \prod_i D[X_i] e^{-S(X_1\dots
X_N)}}{Z_0^N}
\label{ntwo}
\ee
Note that we have not yet specified the state at the edge of the hole
$|z|=\epsilon$.  The choice of this state determines which operator
we actually insert from the chosen twist sector.
The choice of state that gives the lowest dimension operator from the
given twist sector will be the one we will
call $\sigma_n$; other choices of state will give excited states in
the same twist sector.

To specify the state at this edge consider again the operator
$\sigma_n$ inserted at $z=0$. We pass to  local covering
space (which we call
$\Sigma$) by the map which locally looks like
\begin{equation}
{z\approx a t^n}
\label{one}
\end{equation}
where $a$ is  a constant. The $n$ copies of $M$ involved in the twist
give rise to a single copy of $M$ on the cover
$\Sigma$. The hole $|z|<\epsilon$ gives a hole in the $t$ space.  We
glue in a flat disc into
this hole, and extend the path integral perform on
$\Sigma$ to the interior of this disc. This procedure effectively
inserts the identity
operator at the edge of the hole in the $t$ space, and this gives the
lowest dimension
operator $\sigma_n$ in the given twist sector. Note that this procedure
inserts the identity operator with a given
normalization. The twist operator thus constructed will be called
$\sigma_n^\epsilon(0)$; here $\epsilon$ is an
essential regularization, and will cancel out in final expressions
when we compute 3-point functions normalized by
the 2-point functions of the operators.

\subsection{Fermionic variables}

In a supersymmetric theory each copy of $M$ has in addition to the
bosonic coordinates $X_i$ a set of fermionic
coordinates $\psi_i$. Upon insertion of a twist operator these
fermionic variables from different copies of $M$ are
permuted around the twist insertion just like the bosonic variables,
and the correlator of twist operators is defined
in a  manner analogous to (\ref{ntwo}). But we have to take some care
in defining the state at the edge of the hole
$|z|<\epsilon$.  When we pass to the covering space $\Sigma$ by the
map (\ref{one}) the fermionic variables
transform as
\be
\psi_t(t)=\psi_z(z)({dz\over dt})^{1/2}=\psi_z(z)a^{1/2}n^{1/2} t^{n-1\over 2}
\label{three}
\ee
In the $z$ plane we want to have the boundary condition that as we
circle the insertion
point $z=0$ counterclockwise we get
$\psi_1\rightarrow\psi_2\rightarrow\dots\rightarrow\psi_n\rightarrow
\psi_1$. While for the spin zero variables
$X_i$ this meant that we just have $X\rightarrow X$  under transport
around $t=0$ in the $t$ space, we now see that
for the spin
$1/2$ fermionic variables
\be
\psi_t(t)\rightarrow (-1)^{n-1}\psi_t(t)
\label{four}
\ee
under a counterclockwise rotation in the $t$ space around  $t=0$. We now find a
difference between the cases of $n$ odd and $n$ even. The $n$ odd
case is simpler, so we
discuss it first.

\subsection{Twist operators in the supersymmetric theory for odd $n$}

For odd $n$ we see from (\ref{four}) that $\psi_t(t)$ is periodic
around $t=0$, so we can define the state at the edge of
the hole just as in the bosonic case, by gluing in a flat disc to
close the hole and continuing the fields $x, \psi$ to the
interior of the disc. This defines the twist operator
$\sigma_n^\epsilon(0)$, just as in the bosonic case.

This operator however has no charge under the R-symmetry group of the
$N=4$ supersymmetric theory. Thus it is
not a chiral  operator of the theory, since a chiral operator
has $h=j, \bar h=\bar j$ where $h, \bar h$ are the
left and right dimensions of the operator and $j, \bar j$ are the
spins under the left and right $SU(2)$
R-symmetry groups. We therefore seek a natural definition of a chiral
operator that is based on the twist operators
$\sigma_n$.

Each copy of the manifold $M$ gives rise to operators that yield an $N=4$
supersymmetry algebra, for both the homomorphic and the
antiholomorphic variables.
       Let
$J^{k,a}_z(z)$ be the left
$SU(2)$ current of the CFT arising from the $k$th copy of $M$. The
index $a$ takes values 1,2,3. Let
\be
J^+_z= J^1_z+iJ^2_z, ~~~J^-_z= J^1_z-iJ^2_z
\ee

The operator
\be
J^a_z=\sum _{i=1}^N J^{i,a}_z(z)
\label{five}
\ee
       is the diagonal element from the set of $N$  $SU(2)$ currents, and
gives the left $SU(2)$
current of the orbifold CFT.

Now we note that in the presence of the twist operator $\sigma_n(0)$
we can define the operators
\be
J^{+(z)}_{-m/n}\equiv \int {dz\over 2\pi i} \sum _{k=1}^n
~J_z^{k,+}(z) ~e^{-2\pi i
m(k-1)/n} ~z^{-m/n}
\label{six}
\ee
The integral over $z$ is performed over the usual counterclockwise
loop around the origin. The integrand is periodic
around this loop, since $J_z^{k,+}\rightarrow J_z^{k+1,+}$ due to the
cyclic permutation of copies of $M$ around the
twist insertion.  We have called these operators $J^{+(z)}_{-m/n}$ since
they raise the dimension of the twist insertion by
$m/n$. We have included a superscript $(z)$ in these operators to
denote the fact that
they are operators on the $z$ space; this will distinguish them from
modes of the
current operators on the $t$ space which we will consider below.
These operators  raise
the
$SU(2)$ charge under the diagonal
$SU(2)$ by one unit, as can be seen from (\ref{five})
and
\be
J_z^3(z_1) J_z^{k,+}(z_2) \sim {1\over z_1-z_2} J_z^{k,+}(z_2)
\ee

In the $t$ space the operation (\ref{six}) becomes
\be
J^{+(z)}_{-m/n}=\int {dz\over 2\pi i} \sum _{k=1}^n ~J_z^{k,+}(z)
~e^{-2\pi i m(k-1)/n}
~z^{-m/n}\rightarrow \int{ dt\over 2\pi i}  ~J_t^{+}(t)
~a^{-m/n} t^{-m}\equiv a^{-m/n}~J_{-m}^+
\label{seven}
\ee
In the $t$ space the contour runs around a complete counterclockwise
circle around $t=0$. The $n$ different currents $J^{k,+}_z$
in the $z$ plane give rise to the single current $J^+_t$ on the
covering space $\Sigma$, which is the current for the single copy of
$M$ that describes the CFT on $\Sigma$.  To summarize, we find that
the operator $J^{+(z)}_{-m/n}$ in the $z$ space gives the operator
$J^+_{-m}$ in the $t$ space.

Let us return to the construction of the chiral twist operators. We
had observed that the twist operators $\sigma_n$ which
simply permuted copies of $M$ had no charge, and thus had $h>j$ and
were not chiral operators. We now ask if we can stay in
the same twist sector, but add other operators to $\sigma_n$ such
that we increase the charge of the resulting operator. Since
$\sigma_n$ had the minimum dimension in its own twist sector, the
dimension of the operator will also go up in this process. But
if we could achieve $h_{final}=q_{final}$ then we would have
constructed a chiral operator.

The operators $J^{+(z)}_{-m/n}$ allow us to make such a construction. We will
compute the dimensions of our final chiral operators from the 2-point
function in
section 4, but it is helpful to use alternative arguments to compute
the dimension
contributed by various components in the construction of the chiral
operator, and we
will do that below.

Thus
start with
$\sigma_n$, which is just a twist operator that permutes the copies of $M$
around its insertion point. The dimension of $\sigma_n$ is
\be
\Delta_n=\bar\Delta_n={c\over 24}(n-{1\over n})={6\over 24}(n-{1\over
n})={1\over 4}(n-{1\over n})
\label{eight}
\ee
This dimension can be deduced  by looking at the CFT on a cylinder
and noting  that the twist operator changes a theory based
on a set of
$n$ separate copies of $M$ to a theory with one single copy of $M$
but on a spatial section that is $n$ times as long. Thus the
vacuum energy of the ground state changes from $-nc/24 $ to $ -(1/n)
c/24$, and the change gives the dimension of the twist
operator.  If $M$ gives an $N=4$ CFT based on a sigma model with 4
bosons and 4 fermions then we have $c=6$, and (\ref{eight})
follows.

To raise the charge of the operator with minimum increase in
dimension consider the application of $J^+_{-1/n}$. The charge goes
up by one unit, while the dimension increases by only $1/n$. Note
that this low cost in dimension is directly related to the
existence of the twist which allows the fractional dimension charge
operators (\ref{six}); if we did not have a twist then we could
only apply $J^+_{-1}$ which would increase $q$ and $h$ by the same
amount, and so not bring an operator with $h>q$ towards an
operator with $h=q$.

In the $t$ space we have thus applied $J^+_{-1}$ to the identity
operator at $t=0$; thus we just get the state
$J^+_{-1}|0\rangle_{NS}=J_t^+(0)$ where $|0\rangle_{NS}$ is the
Neveu--Schwarz vacuum
in the $t$ space.

We might try to repeat this process with another application of
$J^+_{-1}$ (in the $t$ space), but we find that
$J^+_{-1}J^+_{-1}|0\rangle_{NS}=0$. (This fact and other similar relations
used below can be checked by using the commutation relations
of the current algebra to find the norm of the state, or more simply
by using a bosonic representation of the $N=4$ algebra and
observing that for the manipulations concerned any way of
representing the algebra will yield the same results.) We also find
that
$J^+_{-2}J^+_{-1}|0\rangle_{NS}=0$. Thus the next step is to
construct
$J^+_{-3}J^+_{-1}|0\rangle_{NS}$. We keep proceeding in this way, arriving
at the operator (in the $t$ space)
\be
\sigma^-_n\equiv J^+_{-(n-2)}\dots J^+_{-3}J^+_{-1}|0\rangle_{NS}
\label{ten}
\ee
The dimension of this operator (as seen from the $z$ plane) is
\be
h\equiv \Delta_n^-=\Delta_n+{1\over n}+{3\over n}+\dots +{n-2\over
n}={n-1\over 2}
\label{el}
\ee
The charge is
\be
q={n-1\over 2}=\Delta_n^-
\label{tw}
\ee
and thus the operator is chiral.

The next current operator  that we can apply to (\ref{ten}) is (in
the $t$ space) $J^+_{-n}$.  This raises the charge by one unit, but
the dimension in the $z$ plane also rises by one unit, so the
resulting operator is another chiral operator:
\be
\sigma^+_n\equiv J^+_{-n}J^+_{-(n-2)}\dots J^+_{-3}J^+_{-1}|0\rangle_{NS}
\label{tenq}
\ee
This operator has
\be
h\equiv \Delta_n^+=q={n+1\over 2}
\ee

If we apply the next allowed operator $J^+_{-(n+2)}$ then the charge
goes up by one unit but the dimension goes up by ${n+2\over
n}$; thus we would get $h>q$ and  an operator that is not chiral.

To complete the construction we apply the same steps to the
right moving sector using the current $\bar J^+$. We thus
obtain four chiral operators
\be
\sigma_n^{--}, ~~\sigma_n^{+-}, ~~\sigma_n^{-+}, ~~\sigma_n^{++}
\label{fift}
\ee

It would appear that we could make other operators in this manner,
for example by
replacing
$\dots J^+_{-3} J^+_{-1}|0\rangle_{NS}$  by
$\dots J^+_{-2} J^+_{-2}|0\rangle_{NS}$ or
$\dots G^1_{-3/2}{\tilde G}_{2,-3/2}J_{-1}^+|0\rangle_{NS}$.  But as we
will show in the next section, the operators obtained by the latter
constructions are
proportional to the operators that we have made above, and so no new
operators are
obtained this way. There do exist other chiral operators in the
theory, which use the
details of the structure of the manifold $M$. For example there are
20 $(1,1)$ forms on
$K_3$ which give rise to chiral operators but only 4 $(1,1)$ forms on
$T^4$.  While it
should be possible to make and use these additional chiral operators to compute
correlation functions with our general methods, we have not done so
in this paper. Thus
we will consider only the basic operators (\ref{ten}),  (\ref{tenq}) (and their
counterparts for
$n$ even).

\subsection{Twist operators in the supersymmetric theory for even $n$}

For $n$  even the construction of the chiral primaries from
$\sigma_n$ is slightly different. The operator $\sigma_n$ just
permutes the copies of $M$, so the fermionic variables $\psi_1$  from
the first copy cycle around and return to themselves after
$n$ rotations around the twist insertion in the $z$ plane. But on the
covering surface we see from (\ref{four}) that $\psi_t$
returns to itself after one rotation around $t=0$ but with a change
of sign. This means that we should not close the hole around
$t=0$ by just gluing in a disc and  getting the state $|0\rangle_{NS}$ at
$t=0$. Rather we need to insert an operator that creates a
Ramond vacuum $|0\rangle_R$ at $t=0$, so that the fermion $\psi_t$ will
indeed be antiperiodic around $t=0$.
We will see that this operator must be a spin field ${\cal S}^\alpha,
\alpha=\pm$.

We can extract all the relevant properties of this state  $|0\rangle_R$
without making any reference to the details of the manifold $M$,
just using the fact that $M$  gives rise to an $N=4$ supersymmetric
theory. The Ramond vacuum can be obtained by a spectral flow
starting from the NS vacuum. We list in Appendix A the relevant
formulae for spectral flow. The parameter giving
the amount of  spectral flow is $\eta=1$. In the $t$ space we have
one copy of $M$ so the CFT has $c=6$. The NS vacuum has
$h=q=0$. Then we find that the R vacuum has $h=6/24=1/4, q=-1/2$. We
thus denote
it as $|0^-\rangle_R$. We can act on this state by
$J_0^+$ to obtain another degenerate R vacuum with
$h=1/4, q=1/2$
\be
J_0^+|0^-\rangle_R=|0^+\rangle_R
\ee
The spin fields  ${\cal S^\pm}$  create the states $|0^\pm\rangle_R$
from $|0\rangle_{NS}$.

Starting with $|0^-\rangle_R$ let us make a state with the maximal charge
to dimension ratio, just as we did in the case of $n$ odd above.
The first operator we apply is $J_0^+$. The next lowest dimension
operator that we can
apply is $J_{-2}^+$, and so on. We then find the chiral
       states
\be
\sigma^-_n\equiv J^+_{-(n-2)}\dots J^+_{-2}J^+_{0}|0^-\rangle_{R}
\ee
The dimension and charge of this  operator (as seen from the
$z$ plane) are
\be
h\equiv \Delta_n^-=\Delta_n+{1\over 4n}+{2\over n}+\dots +{n-2\over
n}={n-1\over 2},
~~q={n-1\over 2}=\Delta_n^-
\label{elp}
\ee
Here the contribution $\Delta_n$ arises just as in the case of $n$
odd, and  ${1\over 4n}$ is
the contribution to the dimension in the
$z$ plane coming from the insertion
of the spin field ${\cal S^-}$ in the $t$ plane (this field takes
$|0\rangle_{NS}$ to
$|0^-\rangle_R$).
The fact that the dimension
$1/4$ in the
$t$ plane becomes $1/4n$ in the $z$ plane can be seen from the
    form of
the covering space map $z=a t^n$.

We can apply one more current operator to obtain another  chiral operator:
\be
\sigma^+_n\equiv J^+_{-n}J^+_{-(n-2)}\dots J^+_{-2}J^+_{0}|0^-\rangle_{R}
\ee
This operator has
\be
h\equiv \Delta_n^+={n+1\over 2}, ~~q={n+1\over 2}
\ee

Applying current operators from the right moving sector in an
analogous manner we again obtain four chiral  operators of the
form (\ref{fift}). The charges and dimensions of the operators have
the same form in the case $n$ even as in the case $n$ odd.

\bigskip
{\bf Notation:}\quad  The chiral operators
constructed from $\sigma_n$ are denoted $\sigma_n^{\pm  \pm}$, and their
       dimensions are denoted by $\Delta_n^\pm, \bar\Delta^\pm_n$.   We
will also use the
notation
$\Delta_n={c\over 24}(n-{1\over n})={1\over 4}(n-{1\over n})$.
$\Delta_n$ is the
contribution to the dimension of the chiral operator from the
conformal anomaly.
\bigskip

\subsection{Other members of the chiral operator representation}

We have constructed states that have $h=q$ and are thus chiral
operators of the CFT.
By charge conservation, any correlator of chiral operators will
vanish. To find nonvanishing
correlators we must look at the $SU(2)$ representation of which the
above chiral operator is
the highest weight state $|j,m\rangle=|j,j\rangle$. These other
states have the form
\be
(J_0^-)^k|j, j\rangle = \left(\int \frac{dz}{2\pi i} J^-_z(z)\right)^k|j,
j\rangle
\ee
where $J^-_z$ is  an element of the diagonal $SU(2)$ (\ref{five}).
Thus the operators we
study will be given by applications of $J_0^+$ and $J_0^-$ operators to
the twist operators
$\sigma_n$.

\subsection{Universality in the construction of the chiral primaries}

We note that the construction of the above chiral operators
made no reference to the detailed structure of the manifold
$M$. We used   only the fact that the CFT based on $M$ had $N=4$
supersymmetry. Thus $M$ could be a K3 space at a generic
point in its moduli space, which is not simply an orbifold of a
torus. Thus we are not working with a CFT which can be reduced to a
free field theory. In the computations below we will use a bosonic
representations of current operators to simplify the
calculations; this is allowed because we will be working on a sphere.
But it should be
noted that these computations could all have been performed by using
only the general
properties of the
$N=4$ algebra.

\section{The method of computing correlation functions}
\renewcommand{\theequation}{3.\arabic{equation}}
\setcounter{equation}{0}

In \cite{our} a method was developed to compute the correlation
functions for bosonic orbifolds $M^N/S^N$. We will find that the
bosonic result can be extended to obtain the result in the
supersymmetric case, after we take into account the current operators
that were added to the twist operator $\sigma_n$ to get the
operators $\sigma_n^{--}$ etc. We review here briefly the method
used in \cite{our} and then indicate the way it will be extended to
the supersymmetric
case.

\subsection{The method for bosonic orbifolds}

Let us assume that we have only bosonic variables describing the
sigma model with
target space $M$.  Consider the definition (\ref{ntwo}) of the
correlation function of
twist operators $\sigma_n$. The path integral performed with
twisted boundary conditions becomes a path integral on the covering
space $\Sigma$. The holes at the location of the twist
operators are closed by inserting a disc, and thus $\Sigma$ is a
closed surface. In general $\Sigma$ may have several
disconnected components, but we assume here that there is just one
component, since all different components can be handled in
the same way. The copies of $M$ that are not involved in any of the
twists give a contribution to the partition function that
cancels out between the numerator and denominator of (\ref{ntwo}).

The genus of the surface $\Sigma$ depends on the orders of the twist operators.
At the insertion of the operator $\sigma_{n_j}(z_j)$ the covering
surface $\Sigma$ has a branch point of order $n_j$, which
means that $n_j$ sheets of $\Sigma$ meet at $z_j$. One says that the
ramification order at $z_j$ is $r_j=n_j-1$. Suppose
further that over a generic point $z$ here are $s$ sheets of the
covering surface $\Sigma$. Then the genus $g$ of $\Sigma$
is given by
\be
g=\frac{1}{2}\sum_j r_j - s + 1
\label{tnine}
\ee

The path integral over the $z$ plane with twist insertions becomes a
path integral for a CFT based on only one copy of $M$, on
the surface
$\Sigma$ with no twist insertions or any other operator insertions.
But the metric to be used on $\Sigma$ to
compute the path integral is the metric induced from the $z$ plane,
and it is the dependence of the path integral on this metric
that encodes the dependence of the correlation function on the
location of twist operators in the $z$ plane. We can compute the
path integral for some fiducial metric $\tilde g$ on $\Sigma$,
provided we take into account the correction due to the metric
change by using the conformal anomaly. If $ds^2=e^\phi d\tilde s^2$, then
the partition function $Z^{(s)}$ computed with the metric $ds^2$ is
related to the partition function $Z^{(\tilde s)}$
computed with $d\tilde s^2$ through
\be
Z^{(s)}=e^{S_L}Z^{(\tilde s)}
\ee
where
\be
S_L={c\over 96\pi}\int d^2t \sqrt{-g^{(\tilde s)}}[ \partial_\mu\phi
\partial_\nu\phi g^{(\tilde s)\mu\nu}+2R^{(\tilde s)}\phi]
\label{zfourt}
\ee
is the Liouville action \cite{friedan}.  Here $c$ is the central
charge of the CFT for one copy of $M$.

We can choose the fiducial metric $\tilde g$ to be the flat metric
$dtd\bar t$ everywhere
except for a set of isolated points. Then the metric induced from the
$z$ space is written
as
\be
ds^2=dzd\bar z = e^\phi dtd\bar t.
\label{elOLD}
\ee
Let us call the part of $\Sigma$ that excludes
these isolated curvature points and the punctures arising from the twist
insertions as the `regular region' of $\Sigma$. It was shown in
\cite{our} that for
the cases that we are interested in the contribution to the path integral from
these excluded points is zero, and thus we only have the contribution
$S^{(1)}_L$
from the regular region. Since the metric is flat by construction in
this region we get no
contribution from the term
$R^{(\tilde s)}\phi$ in the action (\ref{zfourt}).
         Thus we have
\be
S_L=S_L^{(1)}={1\over 96\pi}\int d^2t [\partial_\mu\phi \partial^\mu\phi]
\label{sixt}
\ee
We rewrite (\ref{sixt}) as
\be
S_L=-{1\over 96\pi}\int d^2t [ \phi
\partial_\mu\partial^\mu\phi]+{1\over 96\pi}\int_{\partial\Sigma}\phi
\partial_n\phi
\label{sevt}
\ee
Here $\partial_n$ is the normal derivative at the boundaries of
$\Sigma$. From (\ref{elOLD}) we find that
\be
\phi=\log[{dz\over dt}] +\log[{d\bar z\over d\bar t}]
\label{eightt}
\ee
so that
\be
\partial_\mu\partial^\mu\phi=\partial_t\partial_{\bar t}\phi =0
\label{ninet}
\ee
and we get
\be
S_L={1\over 96\pi}\int_{\partial\Sigma}\phi
\partial_n\phi
\label{twenty}
\ee

Thus we get  the desired correlation function of twist insertions on
the $z$ plane as a
sum of contributions from local expressions (\ref{twenty}) from a
finite set of points on
$\Sigma$.  An essential (and generally nontrivial) step in the
calculation is finding the map $t(z)$ that gives the branched cover
of the $z$ plane with the given ramifications at the insertions of
the twist operators.  In this way any correlation function of twist
operators on the $z$ plane for the theory $M^N/S^N$ can be deduced
from a knowledge of the  partition functions $Z_g$ for the
theory for one copy of $M$ on Riemann  surfaces of different genera $g$.

The final primary fields of the CFT are not the twists $\sigma_n$ but
operators $O_n$ which are made from
$\sigma_n=\sigma_{(i_1\dots i_n)}$ by symmetrizing over all different
ways in which the $n$ indices $i_k$ involved in  $\sigma_n$
can be chosen from the set of indices $i=1\dots N$ which denote all
the available copies of $M$. The correlations functions of the
$O_n$ are therefore just given by combinatorial factors multiplying
correlators of the $\sigma_n$.  Note that when we look at
         different choices of indices making up the $O_n$ then we can
get a finite number of different covering surfaces $\Sigma$.
But  we find from the combinatorics that if
$N$ is large then for a given choice of orders of twist operators the
leading contribution comes from the case when
$\Sigma$ is a sphere \cite{our}.  The contribution from the cases
where $\Sigma$
is higher genus is suppressed by a relative factor $1/N^g$.

\subsection{The supersymmetric case}

The above result on combinatorics applies as much to the supersymmetric case as
to the bosonic case. Thus we concentrate on the case where
$\Sigma$ is a sphere in the present paper; this will give us the
leading order result at
large $N$.

      Let us consider the 3-point function of twist operators in the
supersymmetric
theory. As we saw in the previous section, the added complication in the
supersymmetric case is that on $\Sigma$ we should not just close the
puncture at the
location of the twist operator by inserting the identity state
$|0\rangle_{NS}$. Rather, we have to insert current operators at these
locations, and for even $n$ we also need an operator - the `spin
field'-  that takes
the state $|0\rangle_{NS}$ to $|0^-\rangle_{R}$. Thus  $\Sigma$ is found in the
same way as in the case of the bosonic orbifold, but instead of
computing the path integral on
$\Sigma$ we need to compute a correlation function on $\Sigma$.

As in the bosonic case (\ref{none}), let $Z_0$ be the partition
function for the
supersymmetric theory when the target space is one copy of $M$
\be
Z_0=\int _g D[X, \psi] e^{-S[X, \psi]}
\label{znot}
\ee
The partition function is computed in some chosen metric $g$ on the
$z$ space.  (In
\cite{our} the $z$ space was chosen to be a large flat disc  which
was closed to a sphere
by gluing an identical disc at its boundary.)

Choosing the operators to be $\sigma_n^{--}$ for concreteness, we define the
correlation function analogous to (\ref{ntwo}), but with insertions
of currents $J^+, J^-$ required in the construction of the chiral operator:
\bea
&&\langle \sigma_1^{\epsilon --}(z_1)\dots \sigma_k^{\epsilon --}(z_k)\rangle
\nonumber\\
&&\equiv
\frac{1}{Z_0^s}\int_{\rm twisted} \prod_{m=1}^s D[X_m, \psi_m] e^{-S(X_1\dots
X_N, \psi_1\dots \psi_N)}\prod_{i,j}\int\frac{dq_{ij}}{2\pi i}  J^\pm(q_{ij})
(q_{ij}-z_j)^{-n_{ij}/n_j}
\nonumber\\
&&\equiv {{\cal Q}\over Z_0^s}
\label{ntwoq}
\eea
We have assumed that $s$ copies of $M$ are joined by the twisted
boundary conditions,
so that the path integral over the remaining $N-s$ copies of $M$
cancels out between the
numerator and denominator in (\ref{ntwoq}). The $q_{ij}$ are integrated over
contours around the $z_j$, and the integers $n_{ij}$ are given by the
form of the chiral
operators discussed in the last section.

Passing to the covering space
$\Sigma$, which we are assuming to be a sphere,  we get the path
integral for the
theory with one copy of
$M$ but with a metric induced from the $z$ space
\bea
{\cal Q}&=&\int_{g_{induced}} D[X, \psi] e^{-S[X, \psi]}\prod_{i,j}
\int J^\pm_t(q_{ij}) \frac{dq_{ij}}{2\pi i}
(q_{ij}-t_j)^{-n_{ij}}\prod_i {\cal S^-}(t_i)\nonumber \\
&=&e^{S_L^\epsilon}\int_g  D[X, \psi] e^{-S[X,
\psi]}\prod_{i,j} \int J^\pm_t(q_{ij}) \frac{dq_{ij}}{2\pi i}
(q_{ij}-t_j)^{-n_{ij}}\prod_i {\cal S^-}(t_i)\nonumber \\
&=&e^{S_L^\epsilon}(\int_g  D[X, \psi] e^{-S[X,
\psi]})~\langle\prod_{i,j} \int J^\pm_t(q_{ij}) \frac{dq_{ij}}{2\pi i}
(q_{ij}-t_j)^{-n_{ij}}\prod_i {\cal S^-}(t_i)\rangle\nonumber \\
&=&e^{S_L^\epsilon}Z_0~\langle\prod_{i,j} \int J^\pm_t(q_{ij})
\frac{dq_{ij}}{2\pi i}
(q_{ij}-t_j)^{-n_{ij}}\prod_i {\cal S^-}(t_i)\rangle
\label{longeq}
\eea
Here ${\cal S^-}$ are the spin field insertions for
even $n$ operators. $S_L^\epsilon$ is the Liouville
action arising from the conformal anomaly when we rewrite the path
integral for the
metric
$g_{induced}$ (which is induced from the $z$ space onto the $t$
space) in terms of the fiducial metric
$g$ on the sphere that was used to define
$Z_0$ in (\ref{znot}). $S_L^\epsilon$ depends upon the cutoffs used in
defining the twist operators.

The 3-point function normalized by the 2-point functions is
\bea
&&\langle \sigma_1^{--}(z_1) \sigma_2^{ --}(z_2)\sigma_3^{
--\dagger}(z_3)\rangle\nonumber \\
&&\equiv{\langle \sigma_1^{\epsilon --}(z_1) \sigma_2^{\epsilon
--}(z_2)\sigma_3^{\epsilon --\dagger}(z_3)\rangle\over \langle
\sigma_1^{\epsilon --}(0)
\sigma_1^{\epsilon --\dagger}(1)\rangle^{1/2}\langle \sigma_2^{\epsilon --}(0)
\sigma_2^{\epsilon --\dagger}(1)\rangle^{1/2}\langle \sigma_3^{\epsilon --}(0)
\sigma_3^{\epsilon --\dagger}(1)\rangle^{1/2}}\nonumber \\
\label{overall}
\eea
The power of $Z_0$  cancels out in
(\ref{overall}), after we note the relation (\ref{tnine}) and the
fact that a correlator of
the form
$\langle\sigma_n\sigma_n^\dagger\rangle$ is proportional to  $Z_0^{-n}$.
Let us denote the Liouville action for the 3-point function by
$S_L[\sigma_1\sigma_2\sigma_3]$ and for the 2-point functions by
$S_L[\sigma_1\sigma_1^\dagger]$ etc.  Let us also write the
correlator of currents and spin fields in (\ref{longeq}) as
$\langle J,{\cal S}\rangle_{\sigma_1\dots \sigma_k}$. Then we have
\bea
&&\langle \sigma_1^{--}(z_1) \sigma_2^{ --}(z_2)\sigma_3^{
--\dagger}(z_3)\rangle\nonumber \\
&&=e^{S_L[\sigma_1\sigma_2\sigma_3^\dagger]-{1\over 2}
S_L[\sigma_1\sigma_1^\dagger]-{1\over 2}
S_L[\sigma_2\sigma_2^\dagger]-{1\over 2}
S_L[\sigma_3\sigma_3^\dagger]}~{\langle J,{\cal
S}\rangle_{\sigma_1\sigma_2
\sigma_3^\dagger}\over 
\langle J,{\cal S}\rangle_{\sigma_1\sigma_1^\dagger}^{1\over 2}
\langle J,{\cal S}\rangle_{\sigma_2\sigma_2^\dagger}^{1\over
2}\langle J,{\cal S}\rangle_{\sigma_3\sigma_3^\dagger}^{1\over 2}}
\label{longtwo}
\eea

The contribution of the Liouville terms in the above equation was 
shown in \cite{our}
to be the three point function of twist operators in the bosonic 
orbifold theory, where the
central charge was set to $c=6$. (This is discussed in more detail below.)
The computation of 3-point functions in the supersymmetric theory
then just reduces, by (\ref{longtwo}), to a computation of
correlation functions of currents and spin fields on the covering
surface $\Sigma$, for the 3-point function and for the 2-point
functions in the denominator.
The correlation functions of currents
can of course be
computed in any metric on $\Sigma$, since they are independent of the metric.

If all the chiral operators have odd $n$ twist operators then the
only correlation functions that we
need are
      correlators of current operators on $\Sigma$. The other possible
nonzero three point function is
where two of the operators have even $n$ twists. In both these cases
the correlation functions (for
$\Sigma$ a sphere) can be computed purely in terms of the properties
of the chiral algebra of the
$N=4$ supersymmetric theory.  (We can `undo' a contour of $J^+$ from around one
point on $\Sigma$ and replace it by contours surrounding the other
points in the
correlator. The contour moves freely through any other operators $J^+$
on $\Sigma$,
while it picks up a contribution from locations of $J^-$ operators
determined purely by
the $N=4$ algebra.) Given this fact we can use any convenient
representation of this
algebra without losing generality in the choice of
$M$. We use a representation of
current algebra in terms of free bosonic fields;
the current operators are represented by exponentials and
polynomials in these new bosonic variables.  The spin fields are
represented by exponentials in
these bosons as well.

We will normalize the current and spin field insertions in the next
section such that
\be
\langle J,{\cal
S}\rangle_{\sigma_i\sigma_i^\dagger}=1
\label{ltwo}
\ee
  It will also be convenient
to use the notation
\be
\sigma_n^{--}(z)\equiv {\sigma_n^{\epsilon--}(z)\over
\langle\sigma_n^{\epsilon--}(0)\sigma_n^{\epsilon--\dagger}(1)\rangle^
{1/2}}
\label{normalisation}
\ee

It is convenient to compute the fusion coefficients (and thus the 
3-point functions) by
computing
\be
{\langle\sigma_1^{--}(0)\sigma_2^{--}(a)\sigma_3^{--\dagger}(\infty)
\rangle\over
\langle\sigma_3^{--}(0)\sigma_3^{--\dagger}(\infty)\rangle}\equiv 
|C_{1,2,3}^{---}|^2
|a|^{-2(\Delta_1^-+\Delta_2^--\Delta_3^-)}
\label{lone}
\ee

 From (\ref{longtwo}) and (\ref{AppDefCAnom}) we see that the LHS of 
(\ref{lone}) is
\be
{\langle\sigma_1^{--}(0)\sigma_2^{--}(a)\sigma_3^{--\dagger}(\infty)
\rangle\over
\langle\sigma_3^{--}(0)\sigma_3^{--\dagger}(\infty)\rangle}= |C_{1,2,3}|^{12}~
|a|^{-2(\Delta_1+\Delta_2-\Delta_3)}~ {\langle J,{\cal
S}\rangle_{\sigma_1(0)\sigma_2(a)\sigma_3^\dagger(\infty)}\over
\langle J,{\cal
S}\rangle_{\sigma_3(0)\sigma_3^\dagger(\infty) }}
\label{lthree}
\ee
where we have used (\ref{ltwo}). Here $C_{1,2,3}$ is the fusion 
coefficient (computed in
\cite{our}) of twist fields  for a bosonic theory with $c=1$.

To summarize, the three point functions of the
supersymmetric theory will be computed as the product of two
contributions. The first
part is the contribution of the
conformal anomaly, which arises in the map of the
$z$ sphere to the $t$ sphere. This part of the calculation is
identical to that for the bosonic case, after
we note that the central charge for the theory based on one copy of
$M$ is
$c=6$ for the field content of a $N=4$ supersymmetric theory. The
second part is the correlator of
current operators and spin fields on $\Sigma$; these fields will all
be represented by exponentials and
polynomials of bosons on $\Sigma$.

\section{The contribution of the   current insertions}
\label{SectSpecFlow}
\renewcommand{\theequation}{4.\arabic{equation}}
\setcounter{equation}{0}

\subsection{Representing the current algebra by free fields}

As mentioned above when  $\Sigma$  is a sphere we
can use the following simplification. We need only the OPEs of the
chiral algebra
generators to get the correlation function, and so we can represent these
generators in
any manner that reproduces these OPEs.

Let us therefore take a specific example of a system with ${\cal N}=4$
superconformal symmetry, and use it to construct the chiral algebra generators
and their correlation functions on the sphere.
We take two complex  bosons ($X_1$ and $X_2$) and two
complex fermions ($\Psi_1$ and
$\Psi_2$).  The elements of the superconformal algebra for such
system are given by:
\bea
J^a(z)&=&\frac{1}{2}\Psi_1^\dagger\sigma^a\Psi_1+\Psi_2^\dagger\sigma^
a\Psi_2,\qquad
T(z)=\partial X^\dagger_i{\partial}X_i+\frac{1}{2}
\Psi_i^\dagger\partial\Psi_i-\frac{1}{2}\partial\Psi^\dagger_i\Psi_i\\
G^1(z)&=&\sqrt{2}\Psi_2^\dagger\partial X_1-\sqrt{2}\Psi_1\partial X_2,\quad
G^2(z)=\sqrt{2}\Psi_1^\dagger\partial X_1+\sqrt{2}\Psi_2\partial X_2,\\
{\tilde G}_1(z)&=&\sqrt{2}\Psi_2\partial X^\dagger_1-
\sqrt{2}\Psi^\dagger_1\partial X^\dagger_2,\quad
{\tilde G}_2(z)=\sqrt{2}\Psi_1\partial X^\dagger_1+\sqrt{2}\Psi^\dagger_2
\partial X^\dagger_2.
\eea
Let us introduce the real components of the bosonic fields:
\be
X_1=\frac{\phi_1+i\phi_2}{\sqrt{2}},\qquad X_2=\frac{\phi_3+i\phi_4}{\sqrt{2}}
\ee
and bosonize the fermions:
\be
\Psi_1=e^{i\phi_5},\qquad \Psi_2=e^{i\phi_6}.
\ee
One can now rewrite the superconformal generators in terms of six real
bosonic fields $\phi_i$:
\bea
J^3(z)&=&\frac{i}{2}\left(\partial\phi_5-\partial\phi_6\right),\qquad
J^+(z)=\exp(i\phi_5-i\phi_6),\\
J^-(z)&=&\exp(-i\phi_5+i\phi_6)\qquad
T(z)=\frac{1}{2}\sum_{i=1}^{6}\partial \phi^\dagger_i{\partial}\phi_i\\
G^1(z)&=&\exp(-i\phi_6)(\partial\phi_1+i\partial\phi_2)-
\exp(i\phi_5)(\partial\phi_3+i\partial\phi_4),\\
G^2(z)&=&\exp(-i\phi_5)(\partial\phi_1+i\partial\phi_2)+\exp(i\phi_6)
(\partial\phi_3+i\partial\phi_4)
\eea
The components of ${\tilde G}$ can be obtained from $G^a$ by taking a
complex conjugate.

For later use it will be convenient to adopt a notation where all 6
bosons (4 original
bosons and 2 from bosonizing the fermions) are grouped together into a vector
$\phi_a$. Then we write
\bea\label{generatorS}
J^3(z)&=&\frac{i}{2}\sum_{j}e_a\partial_z \phi^a_j(z),\\
J^+(z)&=&\sum_{j}\exp\left(ie_a\phi^a_j(z)\right),\qquad
J^-(z)=\sum_{j}\exp\left(-ie_a\phi^a_j(z)\right),\\
G^1(z)&=&\sum_{j}\left\{\exp\left(-ic_a\phi^a_j(z)\right)
A_b\d\phi^b_j(z)-\exp\left(i d_a\phi^a_j(z)\right)B_b\d\phi^b_j(z)
\right\},\\
G^2(z)&=&\sum_{j}\left\{\exp\left(ic_a\phi^a_j(z)\right)
B_b\d\phi^b_j(z)+\exp\left(-i d_a\phi^a_j(z)\right)A_b\d\phi^b_j(z)\right\}.
\label{generatorF}
\eea
Here we have
\bea
{\bf A}=(1,i,0,0,0,0),\quad {\bf B}=(0,0,1,i,0,0),\nonumber\\
{\bf c}=(0,0,0,0,0,1),\quad {\bf d}=(0,0,0,0,1,0)
\eea
Starting from the general form (\ref{generatorS})--(\ref{generatorF}) and
requiring that the elements of the chiral algebra satisfy their OPEs,
we get the
following constraints on real vectors ${\bf c},\ {\bf d},\ {\bf e}$
and complex vectors ${\bf A},\ {\bf B}$:
\be
{\bf e}\cdot{\bf e}=2,\qquad {\bf c}\cdot{\bf e}=-1,\qquad {\bf
d}\cdot{\bf e}=1,\qquad
{\bf A}\cdot{\bf A}^*={\bf B}\cdot{\bf B}^*=2,
\ee
$$
{\bf c}\cdot{\bf d}={\bf e}\cdot{\bf A}={\bf e}\cdot{\bf B}={\bf
c}\cdot{\bf A}={\bf
c}\cdot{\bf B}= {\bf d}\cdot{\bf A}=
{\bf d}\cdot{\bf B}={\bf A}\cdot{\bf B}={\bf A}\cdot{\bf A}={\bf
B}\cdot{\bf B}=
{\bf A}^*\cdot{\bf B}=0.
$$
\be
{\bf e}={\bf d}-{\bf c},\qquad {\bf d}\cdot{\bf d}={\bf c}\cdot{\bf c}=1.
\ee
These are the only properties that we will explicitly use.

It will  be convenient not to use vectors ${\bf d}$ and ${\bf c}$,
but use instead ${\bf e}$ and
\be
{\bf f}={\bf c}+{\bf d}: \qquad {\bf f}\cdot{\bf f}=2, \qquad {\bf
f}\cdot{\bf e}=0
\ee

In this representation through free bosons it is easy to establish
the claim made in
the previous section about the possible chiral operators in the CFT.
For odd $n$, the
operators $\sigma_n^-$ had the form $J^+_{n-2}\dots
J^+_{-3}J^+_{-1}|0\rangle_{NS}$. The
same dimension and charge can be obtained by replacing $J^+_{-3} J^+_{-1}$ by
$J^+_{-2} J^+_{-2}$ or by $G^1_{-3/2} {\tilde G}_{2,-3/2}J^+_{-1}$. But as we
argue now, these
other operators will all be proportional to the original operator,
and so there will be  no
further chiral operators with this charge and dimension in the twist sector of
$\sigma_n$.

This
operator has charge ${n-1\over 2}$ under $J^3$ and a dimension
${n-1\over 2}$. If we
take any combination of the generators $T, G^\alpha, {\tilde G}^\alpha
J^a$ from
(\ref{generatorS})--(\ref{generatorF}) then we will get expressions which are
of the form `exponential in
the $\phi_a$' times  `polynomial in the
$\phi_a$'.  But the required dimension and charge are obtained only
by the expression
\be\label{stateP}
|p\rangle=b^{-p^2/n}:\exp\left(ipe_a\Phi^a(0)\right):\ ,\qquad p=\frac{n-1}{2}.
\ee
Here $z\approx bt^n$ is the map taking the operator defines in the
$z$ plane to its image in the $t$
space. We will deduce the power of $b$ in (\ref{stateP}) below, but
first let us look at the  charge and
dimension of the operator.  The charge determines the exponential,
and adding any polynomial or
any exponential in a direction orthogonal to the vector
$e_a$ increases  the dimension
without increasing the charge.  This establishes the above claim that
there is a unique
operator with the desired quantum numbers. Note that using the free boson
representation does not limit us to working with a free CFT; the same
result could be
proven by computing the determinant of the matrix of  dot products between the
above mentioned states, and finding that  the matrix has rank 1 after
use of the
commutation relations of the chiral algebra.

Let us now obtain (\ref{stateP}) directly from the definition of the
chiral operator in the $z$ plane,
and thus obtain the required power of $b$. Let us take $n$ odd. We
have from (\ref{seven})
\be
J^{+(z)}_{-1/n}\sigma_n(0)=b^{-1/n} \int {dt\over 2\pi i}J^+_t(t)
t^{-1}=b^{-1/n}J_t^+(0)=b^{-1/n}\exp\left(ie_a\Phi^a(0)\right)
\ee
(Here and in what follows we will identify the operator in the $z$
plane with its
representation in the $t$ space, letting it be clear from the notation which
representation we are referring to.)

If we apply another current operator then we get
\be
J^{+(z)}_{-k/n}J^{+(z)}_{-1/n}\sigma_n(0)=b^{-k/n}b^{-1/n}\int
{dt\over 2\pi i} t^{-k}
\exp\left(ie_a\Phi^a(t)\right)\exp\left(ie_a\Phi^a(0)\right)
\ee
      From the OPE
$\exp\left(ie_a\Phi^a(t)\right)\exp\left(ie_a\Phi^a(0)\right)\sim
t^2$ we see the
claim made earlier that the lowest allowed value of $k$ is $3$. For
$k=3$ we get
\be
J^{+(z)}_{-3/n}J^{+(z)}_{-1/n}\sigma_n(0)=b^{-4/n}\exp\left(2ie_a\Phi^
a(0)\right)
\ee
for the operator in the $t$ space.

Proceeding in this manner we find that
\be
J^{+(z)}_{-(n-2)/n}\dots
J^{+(z)}_{-3/n}J^{+(z)}_{-1/n}\sigma_n(0)=b^{-p^2/n}\exp\left(ipe_a
\Phi^a(0)\right), ~~~p={n-1\over
2}
\ee

In a similar manner we can compute other sets of operators that would
yield chiral states. Let
\be
z\approx bt^n(1+\xi t+O(t^2)).
\ee
be the map from $t$ plane to the $z$ plane near point $t=0$. Then
\bea
J^{+(z)}_{-2/n}J^{+(z)}_{-2/n}\sigma_n(0)&=&J^{+(z)}_{-2/n}b^{-2/n}:(i
e_b\d\Phi^b(0)-
\frac{2\xi}{n})\exp\left(ie_a\Phi^a(0)\right):\nonumber\\
&=&2b^{-4/n}:\exp\left(2ie_a\Phi^a(0)\right):
\eea
so we get the same state as we would get from
$J^{+(z)}_{-3/n}J^{+(z)}_{-1/n}\sigma_n(0)$. (We had argued above
that such had to be the case from
a consideration of the charges and dimensions of the chiral
operators.) Similarly we find that
\be
{\tilde G}_{2,-3/2}G^1_{-3/2}J^+_{-1}\sigma_n(0)
\sim b^{-4/n}:\exp\left(2ie_a\Phi^a(0)\right):
\ee

Let us now consider the case of even $n$.  Now we need to insert a spin
field in the $t$ space to obtain
the vacuum $|0^-\rangle_R$ from $|0\rangle_{NS}$.  We can reproduce the charge,
dimension and OPEs of the spin
field by using the  operator $\exp\left(\pm{1\over
2}ie_a\Phi^a(t)\right)$ to represent the
spin fields ${\cal S^\pm}$. Again using the  fact that the chiral
operators have dimension equal to charge, we find that
\be
J^{+(z)}_{-(n-2)/n}\dots
J^{+(z)}_{-2/n}J^{+(z)}_{0}\sigma_n(0)=b^{-p^2/n}\exp\left(ipe_a\Phi^a
(0)\right), ~~~p={n-1\over
2}
\label{pvalue}
\ee
is the unique state giving the chiral operator with $h=q={n-1\over
2}$ in the twist
sector of $\sigma_n$. Similarly, $\sigma_n^+$ is given by
taking in (\ref{pvalue}) the value $p={n+1\over 2}$.

      Note that we will be using the
correlators of at most two spin fields on a sphere, and for such
correlators the
properties of the spin field encoded in this bosonic representation
entails no loss of
generality. In particular this representation is not equivalent to
assuming that we are dealing with a
theory that can be reduced to free bosons.

\subsection{Two point functions.}

We now wish to compute the 2-point functions
of twist operators, for two reasons. First, we extract the dimensions
of the operators from
the 2-point functions. Second, to define the 3-point function we have
to know the normalization of
the operators that go into the computation, and this is done by
choosing the normalizations of the
2-point functions of the operators with their own conjugates.

We have seen that in the representation through free bosons the
chiral operators take, when lifted
to the $t$ sphere, the form
(\ref{stateP})
\be
\label{statePP}
|p\rangle=A b^{-p^2/n}:\exp\left(ipe_a\Phi^a(0)\right):
\ee
where $p=\frac{n\pm 1}{2}$ for $\sigma^{\pm}_n$. While the structure of the
map $z(t)$ fixes the $b$ dependence in the last equation, the overall
normalization constant $A$ has not been determined so far. To do so, one
needs to evaluate the two point function of such exponentials, and we will do
this  below.

Let us consider the operator  $\sigma_n^{\eps --}$. We place this operator
at $z=0$ and its conjugate
$\sigma_n^{\eps --\dagger}$ at $z=a$. Note that if
$\sigma^-_n$ corresponds to twist $(1,\dots,n)$,  and has charge
${n-1\over 2}$, then
$\sigma^{-\dagger}_n$  corresponds to the twist $(n,\dots,1)$
and has a charge $-(n-1)/2$.  Following the method outlined in the
previous section, we go from the $z$
sphere to the $t$ sphere by the map
\cite{our}:
\be\label{2ptMap}
z=\frac{at^n}{t^n-(t-1)^n}
\ee

The path integral in the presence of the chiral fields then becomes,
on the $t$ sphere, a path integral
with the insertion of exponentials, but with no twists. The metric on
the $t$ sphere is to be induced
from the $z$ sphere, but we use a fiducial sphere metric for $t$
while taking into account the change of metric by the conformal anomaly.
Using the method outlined in section 3 (equation (\ref{ntwoq}), 
(\ref{longeq})) we get
\be\label{Comlete2pt}
\langle\sigma^{\eps --}_n(0)\sigma^{\eps --\dagger}_n(a)\rangle~=
Z_0^{1-n}e^{S^\eps_L}\langle\frac{n-1}{2},z=a|\frac{n-1}{2},z=0\rangle,
\ee
where $\langle\frac{n-1}{2},z=a|\frac{n-1}{2},z=0\rangle$ is a correlation
function of two exponentials\footnote{
To make this expression more compact, we use $\Phi$ to represent the
{\it total} bosonic field, while in the rest of this paper $\Phi$
represents only holomorphic part of such field.} (\ref{statePP}):
\bea\label{2ptfunc}
&&\langle\frac{n-1}{2},z=a|\frac{n-1}{2},z=0\rangle=|A|^2\times\nonumber\\
&&\left\langle
|b_0|^{-(n-1)^2/2n}:\exp\left(i\frac{n-1}{2}e_a\Phi^a(0)\right):
|b_1|^{-(n-1)^2/2n}:\exp\left(-i\frac{n-1}{2}e_a\Phi^a(1)\right):
\right\rangle\nonumber \\
&&~=|A|^2|b_0|^{-(n-1)^2/2n}|b_1|^{-(n-1)^2/2n}.
\eea

The values of $b_0$ and $b_1$ in (\ref{2ptfunc}) are determined by
the asymptotic
behavior of (\ref{2ptMap}) near $t=0$ and $t=1$:
\bea
z=a(-1)^{n-1}t^n+O(t^{n+1}),&&\qquad b_0=a;\\
z=a+a(t-1)^n+O((t-1)^{n+1}),&&\qquad b_1=a.
\eea
Taking this into account, we get an expression for the correlator of two
exponents:
\be
\langle\frac{n-1}{2},z=a|\frac{n-1}{2},z=0\rangle=
|A|^2|a|^{-(n-1)^2/n}.
\ee
Note that in (\ref{longtwo}) the following shorthand notation has been used
for the same two point function:
\be
\langle J,{\cal S}\rangle_{\sigma^{--}_n\sigma_n^{--\dagger}}\equiv
\langle\frac{n-1}{2},z=1|\frac{n-1}{2},z=0\rangle.
\label{shorthand}
\ee
Let us now choose
\be
A=1, ~~~|p\rangle= b^{-p^2/n}:\exp\left(ipe_a\Phi^a(0)\right):
\label{avalue}
\ee
In this case the  expression (\ref{shorthand}) equals unity, and we 
also get unity for the
denominator in (\ref{longtwo}).

Let us now go back to the complete two point function (\ref{Comlete2pt}). The
contribution of conformal anomaly was evaluated in \cite{our} and it has the
following form:
\be
Z_0^{1-n}e^{S^\eps_L}=B_\eps |a|^{-4\Delta_n}.
\ee
Here $B_\eps$ is a coefficient which depends on regularization parameters, but
not on $a$.

Then for the two point function we finally get:
\be
\langle\sigma^{\eps --}_n(0)\sigma^{\eps --\dagger}_n(a)\rangle=
B_\eps |a|^{-4\Delta_n}
|a|^{-(n-1)^2/n}=B_\eps |a|^{-2(n-1)}
\ee
Thus the dimension of the operator $\sigma_n^{--}$ is $\Delta_n^-=\bar
\Delta_n^-={n-1\over 2}$.

The dimensions of the other operators $\sigma_n^{+-}, \sigma_n^{-+},
\sigma_n^{++}$ can be computed
in a similar manner.

\section{Three point function of the form $\langle \sigma^{\pm}_n\sigma^{\pm}_m
\left(\sigma^{\pm}_{m+n-1}\right)^\dagger\rangle$}
\renewcommand{\theequation}{5.\arabic{equation}}
\setcounter{equation}{0}

To start with we will look at a special class of 3-point functions:
\be
\langle \sigma^{\pm}_n(0)\sigma^{\pm}_m(a)
\left(\sigma^{\pm}_{m+n-1}\right)^\dagger(z)\rangle.
\ee
The fact that we have written only one superscript $\pm$ on each $\sigma_n$
indicates that we are looking at only the holomorphic part of the
3-point function. The
$\dagger$ operation changes a permutation
$(1,2,\dots n)$ to the inverse element of the permutation group $(n,
n-1, \dots 1)$, and
also changes the sign of the $J_3$ charge of the operator. If the
permutations of order
$n$ and order $m$
combine into a permutation with a single nontrivial cycle, then the
maximal order of
this cycle is $n+m-1$, and this is the order of permutation chosen
for the third
operator in the above expression. The value of this subclass
of correlators was computed by a recursion argument in
\cite{jevicki}, but we reproduce the result by our method as a warmup
towards the
general case which we study in the next section.

Note that $\sigma_n^-$ has charge ${n-1\over 2}$ while $\sigma_n^+$ has charge
${n+1\over 2}$. Thus by   charge conservation
\bea
\langle \sigma^{+}_n(0)\sigma^{+}_m(a)
\left(\sigma^{+}_{m+n-1}\right)^\dagger(z)\rangle&=&0\nonumber\cr
\langle \sigma^{+}_n(0)\sigma^{-}_m(a)
\left(\sigma^{-}_{m+n-1}\right)^\dagger(z)\rangle&=&0\nonumber\cr
\langle \sigma^{-}_n(0)\sigma^{-}_m(a)
\left(\sigma^{+}_{m+n-1}\right)^\dagger(z)\rangle&=&0.
\eea

The nonvanishing fusion coefficients will be denoted as
\bea
\sigma^{-}_n(0)\sigma^{-}_m(a)&=&a^{\Delta^-_{m+n-1}-\Delta^-_n-\Delta^-_m}
C^{---}_{n,m,n+m-1}\sigma^-_{n+m-1}(0),\\
\sigma^{+}_n(0)\sigma^{-}_m(a)&=&a^{\Delta^+_{m+n-1}-\Delta^+_n-\Delta^-_m}
C^{+-+}_{n,m,n+m-1}\sigma^+_{n+m-1}(0),\\
\eea

To compute the fusion coefficients we use (\ref{lthree}).
In the expression (\ref{lthree}) we have the factors coming from the 
conformal anomaly
\be
  |C_{1,2,3}|^{12}~
|a|^{-2(\Delta_1+\Delta_2-\Delta_3)}=
a^{-\frac{1}{4}(1-\frac{1}{n}-\frac{1}{m}+\frac{1}{n+m-1})}(C_{n,m,n+m-1})^6,
\label{PureSigma}
\ee
\bea
&&\log |C_{n,m,m+n-1}|^2=-\frac{1}{12}\left(n+\frac{1}{n}\right)\log n-
\frac{1}{12}\left(m+\frac{1}{m}\right)\log m\\
&&+\frac{1}{12}\left(q{+}\frac{1}{n}{+}\frac{1}{m}-1\right)\log q-
\frac{1}{12}\left(1{+}\frac{1}{q} {-}\frac{1}{n} {-}\frac{1}{m}\right)
\log\left(\frac{(q{-}1)!}{(m{-}1)!(n{-}1)!}\right).\nonumber
\eea
Here $q=m+n-1$.
  The power of $a$ in (\ref{PureSigma}) arises from the
dimensions of the bosonic
twists $\sigma_n$ which are $\Delta_n={c\over 24}(n-{1\over n})$.

The remainder of the contribution to the OPE comes from the
insertions of $J^+$ (and spin fields for even $n$) at the
images  the twist operators on $\Sigma$. We thus have to compute a correlation
function of these elements of the chiral algebra, and since we are on
the sphere, we
lose no generality by using a representation of the currents in terms
of free bosons.
In the previous section we have shown  (eq. (\ref {avalue}) ) that 
the twist operator
$\sigma^{\pm}_n(0)$ gives rise to the following insertion on the $t$ sphere:
\be
|p\rangle =b_0^{-p^2/n}:\exp\left(ipe_a\Phi^a(0)\right):,
\ee
where $p=\frac{n-1}{2}$ for $\sigma^{-}_n$ and $p=\frac{n+1}{2}$ for
$\sigma^{+}_n$. The value of $b_0$  is given by
the leading behavior of the map near $t=0$:
\be
z=b_0t^n+O(t^{n+1}).
\ee

Consider now a second insertion at the point $z=a$ (which maps to $t=1$):
\be
|q\rangle =b_1^{-q^2/m}:\exp\left(iqe_a\Phi^a(1)\right):
\ee
with $q=\frac{m\pm 1}{2}$, depending on the charge of $\sigma^{\pm}_m(a)$.
To evaluate $b_0$, $b_1$ we recall the map used in \cite{our} to go
from the $z$ sphere
to the $t$ sphere. This map was constructed in terms of the Jacobi polynomials:
\be
z=at^nP^{(n,-m)}_{m-1}(1-2t).
\ee
In particular we will need following asymptotic properties of this map:
\bea\label{asymp0}
z&=&a\frac{(m+n-1)!}{n!(m-1)!}t^{n}
        \quad \mbox{near} \quad z=0,\\
\label{asymp1}
z&=&a+a\frac{(m+n-1)!}{m!(n-1)!}(t-1)^{m}
        \quad \mbox{near} \quad z=a,\\
\label{asymp2}
z&=&a\frac{(m+n-2)!}{(m-1)!(n-1)!}t^{m+n-1}
        \quad \mbox{near} \quad z=\infty
\eea

        The
values of $b_0$ and $b_1$ can be seen from (\ref{asymp0}) and
(\ref{asymp1}) to be:
\be
b_0=a\frac{(m+n-1)!}{n!(m-1)!}\qquad
b_1=a\frac{(m+n-1)!}{m!(n-1)!}.
\ee

Then the charge insertions at $t=0$ and $t=1$ give
\bea\label{MergeTwoHighest}
&&b_0^{-p^2/n}:\exp\left(ipe_a\Phi^a(0)\right):
~~~~b_1^{-q^2/m}:\exp\left(iqe_a\Phi^a(1)\right):\\
&&\quad =n^{p^2/n}m^{q^2/m}
\left(\frac{a(m+n-1)!}{(n-1)!(m-1)!}\right)^{-p^2/n-q^2/m}
:\exp\left[ipe^b\Phi^b(0)+iqe^b\Phi^b(1)\right]:\nonumber\\
\eea

To compute the contribution of (\ref{MergeTwoHighest}) to the fusion
coefficient we
recall the way we had computed the fusion coefficients of the bosonic twists in
\cite{our}.  The $z$ plane was cut off at a large radius
$|z|=1/\delta$, and a second disc
was glued in at this boundary to make the $z$ plane into a sphere
with an explicitly
chosen metric. The map $z(t)$ taking multivalued functions on the $z$
sphere to single
valued functions on the covering space thus mapped a closed surface
(a sphere) to a
closed surface $\Sigma$ (which in the present case is also a sphere).
It was important
to explicitly close all surfaces in order to use the argument of the
conformal anomaly
to compute the correlation function.

In the 3-point function we place operators at $z=0$, $z=a$ and $z=\infty$. We
normalized the insertions at $z=0$ and $z=a$. We could also normalize
the insertion at
$z=\infty$ and directly compute the 3-point function, but because the
operator at
infinity is in a different coordinate patch it is easier to proceed
in a slightly different
way. We place an operator $\sigma_q^\infty$ at infinity, without regard to its
normalization, and compute the ratio
\be
{\langle \sigma_n(0)\sigma_m(a)\sigma_q^\infty(\infty)\rangle\over \langle
\sigma_q(0)\sigma_q^\infty(\infty)\rangle}=
C_{mnq}a^{-(\Delta_n+\Delta_m-\Delta_q)}
\label{type}
\ee
(where $\sigma_q$ is correctly normalized) to compute the OPE coefficient.

To extend the same method to the present case we can get identical
insertions at infinity in the case of $\sigma_n(0) \sigma_m(a)$ and the case of
$\sigma_q(0)$ if we choose the map to the covering space for $\sigma_q(0)$ to
agree at large $z$ with the map to the covering space for the
insertion $\sigma_n(0)
\sigma_m(a)$. From (\ref{asymp2}) we see that the map  for
$\sigma_{m+n-1}(0)$ should be chosen to be
\be
z=a\frac{(m+n-2)!}{(m-1)!(n-1)!}t^{m+n-1}\equiv b_\infty t^{m+n-1}
        \quad \mbox{near} \quad z=\infty
\ee
Writing $r=p+q$ for the charge of $\sigma_{n+m-1}^\pm(0)$ we get for  the
normalized charge insertion
\be
|r\rangle =b_\infty^{-r^2/n}:\exp\left(ire_a\Phi^a(0)\right):,
\ee
where
\be
b_{\infty}=a\frac{(m+n-2)!}{(m-1)!(n-1)!}.
\ee

Then we get from the charge insertions the contribution to (\ref{lthree})
\bea
&& {\langle J,{\cal
S}\rangle_{\sigma_1(0)\sigma_2(a)\sigma_3(\infty)}\over
\langle J,{\cal
S}\rangle_{\sigma_3(0)\sigma_3(\infty) }}\nonumber \\
&&=\frac{\langle
b_0^{-p^2/n}:e^{ipe_a\Phi^a(0)}:~~
        b_1^{-q^2/m}:e^{ipe_b\Phi^b(1)}:~~
:e^{-i(p+q)e_b\Phi^b(\infty)}:\rangle}
{\langle b^{-r^2/(m+n-1)}_{\infty}
:e^{i(p+q)e_c\Phi^c(0)}:~~
:e^{-i(p+q)e_d\Phi^d(\infty)}:\rangle}\nonumber \\
&&=b_0^{-p^2/n}
b_1^{-q^2/m}b^{r^2/(m+n-1)}_{\infty}
\label{charges}
\eea

Let us now evaluate $C_{n,m,n+m-1}^{---}$.   In this case:
\be
p=\frac{n-1}{2},\qquad  q=\frac{m-1}{2},\qquad r=\frac{n+m-2}{2}
\ee
Combining the contribution (\ref{charges})  with the contribution
(\ref{PureSigma})
we get
\be
|C^{---}_{n,m,n+m-1}|^2=|C_{n,m,n+m-1}|^{12}
|a|^{-\frac{1}{2}(1-\frac{1}{n}-\frac{1}{m}+\frac{1}{n+m-1})}~|b_0|^{-2p^2/n}
|b_1|^{-2q^2/m}|b_{\infty}|^{2r^2/(m+n-1)}
\ee
where we now write the combined holomorphic and antiholomorphic sector
contributions.  Substituting the value of the bosonic fusion coefficient
(\ref{PureSigma}) and making algebraic simplifications, we get
\be
\sigma_n^{--}(0)\sigma_m^{--}(a)\sim |C^{---}_{n,m,n+m-1}|^2
\sigma_{n+m-1}^{--}(0)
\label{power}
\ee
with
\be
|C^{---}_{n,m,n+m-1}|^2=\frac{m+n-1}{mn}.
\ee
Note that the power of $a$ cancels out in (\ref{power}), reflecting
the fact that in the
supersymmetric theory $\Delta^-_{m+n-1}=\Delta^-_n+\Delta^-_m$.

Similarly one finds
\be\label{COneOver}
|C^{+-+}_{n,m,n+m-1}|^2=\frac{n}{m(m+n-1)}
\ee

\section{General three point function for the sphere.}
\label{Sect3Pt}
\renewcommand{\theequation}{6.\arabic{equation}}
\setcounter{equation}{0}

Let us now consider the general 3-point function $\langle
\sigma_n^\pm\sigma_m^\pm\sigma_q^\pm\rangle$. The twist operators carry
representations of the $su(2)\times su(2)$  symmetry group. Thus
representations of
the first two twist operators have to combine to give the (conjugate) to the
representation of the third twist operator. Since there is at most
one way to combine
two representations of $su(2)$ to a third representation,  we see
that if we can
compute the fusion coefficient for any chosen members of the three
representations,
then the general fusion coefficient can be deduced from this in terms of the
Clebsch-Gordon coefficients.

In the simple case studied in the last section all operators had
charge vectors of the
form $|j,m\rangle=|Q,Q\rangle$ or $|j,m\rangle=|Q,-Q\rangle$, and so
the charge could be
represented as a pure
exponential. The main additional complication in the general case is that
all charge vectors
cannot be taken to have this simple form.  We will let the operator at
$z=0$ have the form $|j,m\rangle=|Q, Q-d\rangle$, while the operator
at $z=a$ has the form
$|j,m\rangle=|P,P\rangle$ and the operator at $z=\infty$ has the form
$|j,m\rangle=|Q-d+P, -(Q-d+P)\rangle$.
Thus the operator at
$z=0$ needs to be modified by the action of lowering operators acting on an
exponential, and we carry out this modification below. The other steps in the
calculation parallel those of the previous section.

\subsection{Constructing the state $|Q, Q-d\rangle$}

We will  construct a family of twist operators by acting on the highest
weight state (\ref{statePP}) by the charge operator corresponding to
$J^-$.  Using the fact that $J^-_z$  has holomorphic dimension unity
we  can rewrite
the corresponding charge on the  covering space:
\be
Q^-=\oint\frac{dz}{2\pi i}J^-(z)=
\oint\frac{dz}{2\pi i}\sum_{j}\exp\left(-ie_a \phi^a_j(z)\right)=
\oint\frac{dt}{2\pi i}\exp\left(-ie_a \Phi^a(t)\right).
\ee
Acting by this charge on the exponent (\ref{statePP}), we get:
\be\label{OneQ-In}
\oint\frac{dz}{2\pi i}J^-(z)|Q\rangle=
b^{-Q^2/n}\oint_t\frac{dq}{2\pi i}
(q-t)^{-2Q}:\exp\left(iQe_a\Phi^a(t)-ie_a\Phi^a(q)\right):.
\ee
For the BPS operators constructed in section \ref{SectSpecFlow}, the product
$2Q$ is an integer, so one can evaluate the integral
in (\ref{OneQ-In}):
\be
Q^-:\exp\left(iQe_a\Phi^a(t)\right):=\left.\frac{1}{(2Q-1)!}\partial_q^{2Q-1}
:\exp\left(iQe_a\Phi^a(t)-ie_a\Phi^a(q)\right):\right|_{q=t}
\ee
Note that the action of $Q^-$ does not change the conformal dimension of the
operator, since we get a polynomial of order $2Q-1$ multiplying the
exponential, and
$Q^2=(Q-1)^2+(2Q-1)$.

For the double action of the operator $Q^-$ on BPS state one then gets:
\bea
&&(Q^-)^2:\exp\left(iQe_a\Phi^a(t)\right):\\
&&=\oint \frac{dq_2}{2\pi i}
:\exp\left(-ie_a\Phi^a(q_2)\right):\lim_{q_1\rightarrow t}
\frac{1}{(2Q{-}1)!}\partial_{q_1}^{2Q{-}1}
:\exp\left(iQe_a\Phi^a(t){-}ie_a\Phi^a(q_1)\right):\nonumber\\
&&=\lim_{q_1\rightarrow t}\frac{1}{(2Q{-}1)!}\partial_{q_1}^{2Q{-}1}
\oint \frac{dq_2}{2\pi i}
\frac{(q_2{-}q_1)^{2}}{(q_2{-}t)^{2Q}}
:\exp\left(iQe_a\Phi^a(t){-}ie_a\Phi^a(q_1){-}ie_a\Phi^a(q_2)\right):
\nonumber\\
&&=\lim_{q_1\rightarrow t}\lim_{q_2\rightarrow t}
\left(\frac{1}{(2Q{-}1)!}\partial_{q_1}^{2Q{-}1}\right)
\left(\frac{1}{(2Q{-}1)!}\partial_{q_2}^{2Q{-}1}\right)
\nonumber\\
\nonumber\\
&&\qquad \times (q_2{-}q_1)^2
:\exp\left(iQe_a\Phi^a(t){-}ie_a\Phi^a(q_1){-}ie_a\Phi^a(q_2)\right):\nonumber
\eea

One can generalize the above formula using induction:
\bea\label{GenThreeState}
&&(Q^-)^d~:\exp\left(iQe_a\Phi^a(t)\right):~=~\lim_{q_1\rightarrow t}\dots
\lim_{q_d\rightarrow t}
\prod_{l=1}^{d}\left(\frac{1}{(2Q{-}1)!}\partial_{q_l}^{2Q{-}1}\right)\\
&&\qquad\times\prod_{i<j}^{d}(q_j-q_i)^2
~:\exp\left(iQe_a\Phi^a(t)-\sum_{l=1}^{d}ie_a\Phi^a(q_l)\right):\nonumber
\eea

Let us evaluate the norm of the state (\ref{GenThreeState}). To do this we
only need to use the structure of $SU(2)$ algebra:
\be
[Q^+,Q^-]=2Q^3,\qquad [Q^3,Q^\pm]=\pm Q^\pm.
\ee
One can write the two point function of twist operators in terms of the norms
of appropriate representations of $SU(2)$:
\be
\langle \left(\left(Q^-\right)^d\sigma^\pm_{n}(0)\right)
\left(Q^+\right)^d\sigma^{\pm\dagger}_{n}(w)\rangle=
\|\left(Q^-\right)^m|\frac{n\pm 1}{2},\frac{n\pm 1}{2}\rangle\|^2.
\ee
Then standard manipulations give:
\bea
&&\langle Q,Q|\left(Q^+\right)^d\left(Q^-\right)^m|Q,Q\rangle
=2\sum_{j=0}^{d-1}\left(Q-j\right)
\langle Q,Q|\left(Q^+\right)^{d-1}
\left(Q^-\right)^{d-1}|Q,Q\rangle\nonumber\\
&&\qquad=d(2Q+1-d)\|\left(Q^-\right)^{d-1}|Q,Q\rangle\|^2,
\nonumber\\
&&\|\left(Q^-\right)^d|Q,Q\rangle\|=
\left(\frac{d!(2Q)!}{(2Q-d)!}\right)^{1/2}
\||Q,Q\rangle\|.
\eea

Thus we can define the normalized state:
\be
|Q,Q-d\rangle=\left(\frac{(2Q-d)!}{d!(2Q)!}\right)^{1/2}
\oint \frac{dp_1}{2\pi i}J^-(p_1)\dots \oint \frac{dp_d}{2\pi i}J^-(p_d)
|Q,Q\rangle.
\ee

In the bosonised formulation we get:
\bea\label{NormGener}
&&|Q,Q-d\rangle=\left(\frac{(2Q-d)!}{d!(2Q)!}\right)^{1/2}b^{-Q^2/n}
\lim_{q_1\rightarrow t}\dots
\lim_{q_d\rightarrow t}
\prod_{l=1}^{d}\left(\frac{1}{(2Q{-}1)!}\partial_{q_l}^{2Q{-}1}\right)\\
&&\qquad\times\prod_{i<j}^{d}(q_j-q_i)^2
:\exp\left(iQe_a\Phi^a(t)-\sum_{l=1}^{d}ie_a\Phi^a(q_l)\right):\nonumber
\eea

\subsection{The contribution of the current insertions}

     Let us work
out the
contribution from the current insertions on the $t$ sphere.  This 
contribution corresponds to
the factor in (\ref{lthree})
\be
{\langle J,{\cal
S}\rangle_{\sigma_1(0)\sigma_2(a)\sigma_3(\infty)}\over
\langle J,{\cal
S}\rangle_{\sigma_3(0)\sigma_3(\infty) }}
\ee

We
are looking for the
fusion coefficient
        \be
(Q,Q-d_2)\times (P,P)
\rightarrow (Q+P-d,Q+P-d),
\ee
one should consider the fusion rule for the following CFT operators:
\be
|Q,Q-d\rangle_0\ |P,P\rangle_a\approx
{\cal A}|Q+P-d,Q+P-d\rangle_0.
\ee

We first note that
\bea
&&\left\{(Q^-)^{d}:\exp\left(iQe_a\Phi^a(0)\right):\right\}
:\exp\left(iPe_a\Phi^a(1)\right):\\
&&\qquad\approx \lim_{q_i\rightarrow 0}\prod_{l=1}^{d}\left(\frac{1}{(2Q-1)!}
\partial_{q_l}^{2Q-1}\right)\prod_{i<j}^{d}(q_j-q_i)^2\prod_{k=1}^{d}
(1-q_k)^{-2P}\nonumber\\
&&\qquad\times:\exp\left(i(P+Q-d)e_a\Phi^a(0)\right)[1+\dots]:
\eea
where the $\dots$ in the last term indicate polynomials in $\partial\Phi_a(0),
\partial^2\Phi_a(0)$ etc.  We are again using a computation analogous to
(\ref{type}),  and will thus  be computing the correlator of the 
above expression with a
pure exponential operator at infinity. In this situation the terms 
represented by $\dots$ in
the above expression will give a contribution that is vanishing 
compared to the first term, and
can thus be ignored. Taking into account the normalization 
(\ref{NormGener}), we finally
get:
\bea\label{DefCalA}
&&{\cal A}=b_0^{-Q^2/n}b_1^{-P^2/n}b_\infty^{S^2/q}
\left(\frac{(2Q-d)!}{d!(2Q)!}\right)^{1/2}\\
&&\qquad\times
\lim_{q_i\rightarrow 0}\prod_{l=1}^{d}\left(\frac{1}{(2Q{-}1)!}
\partial_{q_l}^{2Q-1}\right)\prod_{i<j}^{d}(q_j-q_i)^2\prod_{k=1}^{d}
(1-q_k)^{-2P}\nonumber
\eea
The value of $S$ in the above expression is  $S=\frac{q\pm 1}{2}$ for
$\sigma_q^\pm$.

\bigskip
{\bf Notation:} \quad It will be helpful in what follows to introduce
the following
notation. We  will characterize the twist operator $\sigma^\pm_n$ by two
numbers: $n$ and
$1_n$, where $1_n=1$ for $\sigma^+_n$ and $1_n=-1$ for $\sigma^-_n$. We also
introduce the notation
\be
\sigma^{1_n(d)}_n
\ee
for twist operator corresponding to the state (\ref{NormGener}). Here
the value of $Q$
is given by
$Q=\frac{n+1_n}{2}$ and $d$ is the number of lowering operators $Q^-$
that have been
applied to the state $|Q, Q\rangle$.
\bigskip

To evaluate the expression ${\cal A}$ we need to use the map
which takes the $z$ sphere to the $t$ sphere.
In \cite{our} we have shown that such a map is unique up to $SL(2,C)$
transformation and its explicit form is given in the appendix \ref{AppBoson}.
Here we will need only some properties of this map, namely its
behavior near the
ramification points:
\bea\label{asym1}
&&t\rightarrow 0:\qquad z\approx a
\frac{d_1!d_2!(n-d_2-1)!}{n!(d_1-n)!(n-1)!}t^n,\\
\label{asym2}
&&t\rightarrow 1:\qquad z\approx a+a
\frac{d_1!d_2!(m-d_2-1)!}{m!(d_1-m)!(m-1)!}(t-1)^m,\\
\label{asym3}
&&t\rightarrow \infty:\qquad z\approx
a\frac{d_2!(d_1-d_2-1)!(d-1-d_2)!}{d_1!(d_1-n)!(d_1-m)!}t^{d_1-d_2}.
\eea

Using the above information, we get:
\bea
&&\log{\cal A}=\left(-\frac{(n+1_n)^2}{4n}-\frac{(m+1_m)^2}{4m}-
\frac{(q+1_q)^2}{4q}\right)\log d_1!
\nonumber\\
&&+\left\{\left(\frac{(n{+}1_n)^2}{4n}-\frac{(m{+}1_m)^2}{4m}-
\frac{(q{+}1_q)^2}{4q}\right)
\log(d_1{-}n)!+\left(n\leftrightarrow m\right)+
\left(n\leftrightarrow q\right)\right\}\nonumber\\
&&+\left\{\frac{(n+1_n)^2}{4n}\log\left(n!(n-1)!\right)+
\left(n\leftrightarrow m\right)+
\left(n\leftrightarrow q\right)\right\}\nonumber\\
&&+\log\left\{
\lim_{q_i\rightarrow 0}\prod_{l=1}^{d}\left(\frac{1}{(n+1_n-1)!}
\partial_{q_l}^{n+1_n-1}\right)\prod_{i<j}^{d}(q_j-q_i)^2\prod_{k=1}^{d}
(1-q_k)^{-m-1_m}
\right\}\nonumber\\
&&-\frac{1}{2}\log\left(\frac{{d}!(n+1_n)!}{(n+1_n-d)!}\right)+
\left(\frac{S^2}{q}-\frac{P^2}{n}-\frac{Q^2}{m}\right)\log a,
\eea
where
\be
d=d_2+\frac{1}{2}(1_n+1_m-1_q+1).
\ee

\subsection{The fusion coefficient $ C^{---}_{n,m,q}$}

Collecting the contributions from the conformal anomaly and charge 
insertions, one finds the
following OPE
\bea
\sigma^{1_n(d)}_n(0)\sigma^{1_m}_m(a)&\approx&
a^{\Delta^{1_q}_q-\Delta^{1_n}_n-\Delta^{1_m}_m}
C_{n,m,q}^{1_n(d),1_m,1_q}
\sigma^{1_q}_{q}(0),\\
&&\nonumber\\
C_{n,m,q}^{1_n(d),1_m,1_q}&=&
(C_{n,m,q})^{6} {\cal A} a^{-S^2/q+P^2/n+Q^2/m}
\eea
Here we are writing only the holomorphic part of the OPE; the
complete OPE will have
representations of the two $su(2)$ factors and will be constructed at the end.

Taking into account the expression for fusion coefficient
of the bosonic  twists found in \cite{our} (see appendix \ref{AppBoson}), we
finally get an
expression for the holomorphic part of the fusion coefficient
\bea\label{theResult}
&&\log C_{n,m,q}^{1_n(d), 1_m, 1_q}=
(-d_2+2)\log d_1!+d_2\left(\log(d_1-n)!+\log(d_1-m)!\right)
\\
&&\nonumber\\
&&\qquad+(d_2-2)\log d_2!-2\log(q-1)!-\frac{1}{2}\log(mnq^3)+d_2(d_2-1)\log 2
+\log{\cal D}\nonumber\\
&&\qquad+\log L^d_{n+1_n,m+1_m}-
\frac{1}{2}\log\left(\frac{d!(n+1_n)!}{(n+1_n-d)!}\right)
-\frac{1}{2}(3+1_n+1_m+1_q)\log d_1!+\nonumber\\
&&\qquad\left\{
\frac{1_n{-}1_m{-}1_q{-}1}{2}\log(d_1-n)!+\frac{1{+}1_n}{2}
\log\left(n!(n-1)!\right)+
(n\leftrightarrow m)+(n\leftrightarrow q)
\right\}.\nonumber
\eea
Here we introduced the notation:
\be\label{PostuLate}
L^d_{n,m}=\left\{
\lim_{q_i\rightarrow 0}\prod_{l=1}^{d}\left(\frac{1}{(n-1)!}
\partial_{q_l}^{n-1}\right)\prod_{i<j}^{d}(q_j-q_i)^2\prod_{k=1}^{d}
(1-q_k)^{-m}
\right\}.
\ee
To get a contribution form antiholomorphic sector, one should replace $1_n$,
$1_m$ and $1_q$ in (\ref{theResult}) by corresponding value for
antiholomorphic part (${\bar 1}_n$, ${\bar 1}_m$ or ${\bar 1}_q$).

The expression (\ref{theResult}) looks complicated for two reasons.
Firstly  it
contains the discriminant ${\cal D}$ which is known only as a finite product
(\ref{discriminant}). Secondly there is  the  term $L^d_{n,m}$ defined through
derivatives and limits.

First we note that (\ref{theResult}) simplifies enormously for the
low values of $d_2$.
In particular, one can easily evaluate $ C^{---}_{n,m,q}$ for $d_2=0$ and
$d_2=1$:
\bea
C_{n,m,m+n-1}^{-,-,-}&=&
\left(\frac{m+n-1}{mn}\right)^{1/2},\\
C_{n,m,m+n-3}^{-(1),-,-}&=&
\left(\frac{(m+n-2)^2}{mn(m+n-3)}\frac{(n-2)!}{(n-1)!}\right)^{1/2}.
\eea
The first of these two expressions is the result (\ref{COneOver}) derived for a
special case in the previous section.  Investigating a few further
cases, and using the  symmetry properties of $C$
lead us  to guess that the complicated expression (\ref{theResult}) for the
case of three operators $\sigma^-$ is equal to the following simple expression:
\be\label{theRes1}
C_{n,m,q}^{-(d_2),-,-}=
\left(\frac{d_1^2(d_2!)^2}{mnq}\frac{(n-d_2-1)!}{d_2!(n-1)!}\right)^{1/2}.
\ee

While we could not prove the  agreement of (\ref{theResult}) and
(\ref{theRes1})
analytically, we have verified it for $d_2\le 5$ and arbitrary
values of $n$ and $m$ using a symbolic manipulations program (Mathematica).
Assuming the equality of these two forms of the result, one arrives
at the following
expression for the $L^d_{n,m}$:
\bea
L^d_{n,m}&=&(d!)^3\left[\frac{(n+m+1-d)!}{d!(n-d)!(m-d)!}\right]^d
\left[\frac{(n+m+1-2d)!}{(n+m+1-d)!}\right]^2\frac{n+m+1-d}{n+m+1-2d}\\
&\times&\prod_{j=1}^{d}j^{2d-2-j}(j-n-1)^{1-j}(j-m-1)^{1-j}(j-n-m-2+d)^{j-d}
\label{expression}
\eea

\subsection{The fusion coefficients for general elements of the
representations}

${}$ From now on we will assume that (\ref{expression})  is correct,
and use this
to write compact  expressions for  general fusion coefficients.

We can rewrite (\ref{theRes1}) in the more symmetric form:
\be\label{theRes2}
C_{n,m,q}^{-(d_2),-,-}=
\left(\frac{d_1^2}{mnq}\frac{[(n+m-q-1)/2]![(n-m+q-1)/2]!}{(n-1)!}\right)^{1/2}
\ee
Since we are considering the fusion of the different representations of
$SU(2)$, we anticipate the appearance of the appropriate $3j$ symbols in the
final answer. In terms of the standard notation $|j,m\rangle$ for the
representations of
$SU(2)$, the operator $\sigma^-_n(0)$ can be written as
\be
|\frac{n-1}{2},\frac{n-1}{2}-d_2\rangle,
\ee
so the case computed in the above subsection  corresponds to the
following
$3j$ symbol:
\be
\left(
\begin{array}{ccc}
\frac{n-1}{2}&\frac{m-1}{2}&\frac{q-1}{2}\\
\frac{q-m}{2}&\frac{m-1}{2}&\frac{1-q}{2}
\end{array}
\right).
\ee
One can easily evaluate this particular type of the $3j$ symbol (see, for
example \cite{landau}):
\be
\left(
\begin{array}{ccc}
\frac{n-1}{2}&\frac{m-1}{2}&\frac{q-1}{2}\\
\frac{q-m}{2}&\frac{m-1}{2}&\frac{1-q}{2}
\label{coeff}
\end{array}
\right)=\left(\frac{(m-1)!(q-1)!}{d_1![(m+q-n-1)/2]!}\right)^{1/2}.
\ee
We are to take the positive sign of the square root here and in all
similar expressions in
what follows. The fact that the coefficients (\ref{coeff}) are
positive allow us to take such
square roots and thus write separate factors for the holomorphic and
aniholomorphic
fusion coefficients, without introducing any extra phase factors.

      Using the above
expression, we can rewrite (\ref{theRes2}) in the final form:
\bea
&&C_{n,m,q}^{-(d_2),-,-}=
\left(\frac{d_1^2}{mnq}
\frac{d_1!~[\frac{n+m-q-1}{2}]!~[\frac{n+q-m-1}{2}]!~[\frac{m+q-n-1}{2}]!}
{(n-1)!~(m-1)!~(q-1)!}\right)^{1/2}\nonumber\\
&&\qquad\times\left(
\begin{array}{ccc}
\frac{n-1}{2}&\frac{m-1}{2}&\frac{q-1}{2}\\
\frac{q-m}{2}&\frac{m-1}{2}&\frac{1-q}{2}
\end{array}
\right).
\eea
In particular, it is convenient to define the reduced fusion coefficient,
which describes a fusion of different representations of $SU(2)$ and does
not specify the ``orientation of the spins''.
\be
{\hat C}_{n,m,q}^{---}=\left(
\frac{d_1^2}{mnq}
\frac{d_1!~[\frac{n+m-q-1}{2}]!~[\frac{n+q-m-1}{2}]!~[\frac{m+q-n-1}{2}]!}
{(n-1)!~(m-1)!~(q-1)!}\right)^{1/2}
\ee

Then the fusion rule for the specific example of twist operators $\sigma^{--}$
reads:
\bea
\sigma_n^{--(s_1,{\bar s}_1)}(0)&&{\hskip -15pt}\sigma_m^{--(s_2,{\bar s}_2)}
(a)\sim
|a|^{-(2\Delta_n+2\Delta_m-2\Delta_q)}\sigma_q^{--(s_3,{\bar s}_3)}(0)\nonumber\\
&&\times|{\hat C}_{n,m,q}^{---}|^2
\left(
\begin{array}{ccc}
\frac{n-1}{2}&\frac{m-1}{2}&\frac{q-1}{2}\\
s_1&s_2&s_3
\end{array}
\right)
\left(
\begin{array}{ccc}
\frac{n-1}{2}&\frac{m-1}{2}&\frac{q-1}{2}\\
{\bar s}_1&{\bar s}_2&{\bar s}_3
\end{array}
\right)
\eea

\subsection{The fusion coefficients $C^{1_n,1_m,1_q}_{m,n,q}$}

We have also analyzed the general  fusion coefficients
$C^{1_n,1_m,1_q}_{m,n,q}$. We present the computation of another
case, the coefficient
$C^{+-+}_{m,n,q}$,  in  appendix C.  The final result for the
reduced fusion coefficients is:
\be\label{TheFinal}
{\hat C}_{n,m,q}^{1_n,1_m,1_q}=
\left(\frac{\left(1_n n+1_m m+1_q q+1\right)^2}{4mnq}
\frac{\Sigma!\ \alpha_n!\alpha_m!\alpha_q!}
{(n+1_n)!~(m+1_m)!~(q+1_q)!}\right)^{1/2}
\ee
where
\bea
\Sigma&=&\frac{1}{2}\left(n+1_n+m+1_m+q+1_q\right)+1,\\
\alpha_n&=&\Sigma-n-1_n-1
\eea
Note that the parameters $1_n, 1_m, 1_q$ can be chosen independently for the
holomorphic and antihlomorphic parts of the twist operator. The full
fusion coefficient
is then a product of (\ref{TheFinal}) from the left and right sides,
together with the
Clebsch-Gordon coefficients from the left and right $su(2)$ representations.

\subsection{Combinatoric factors and large N limit.}

The twist operators we have considered so far do not represent proper fields
in the conformal field theory. In the orbifold CFT there is one twist field
for each conjugacy class of the permutation group, not for each
element of the group \cite{vafaorb}. The true CFT
operators that represent the twist fields can be constructed by
summing over the group orbit, for example:
\be
O^{1_n{\bar 1}_n}_n=\frac{\lambda_n}{N!}\sum_{h\in G}
\sigma^{1_n{\bar 1}_n}_{h(1\dots n)h^{-1}}\ .
\ee
Here $G$ is the permutation group $S_N$ and the normalization constant
${\lambda_n}$ can be determined from the normalization condition. Namely if
one starts from normalized $\sigma$ operators:
\be
\langle\sigma^{1_n{\bar 1}_n}_n(0)\sigma^{1_n{\bar 1}_n\dagger}_n(1)\rangle=1,
\ee
then for $O$ we get:
\be
\langle O^{1_n{\bar 1}_n}_n(0)O^{1_n{\bar 1}_n\dagger}_n(1)\rangle=
\lambda_n^2n\frac{(N-n)!}{N!}
\langle\sigma^{1_n{\bar 1}_n}_n(0)\sigma^{1_n{\bar 1}_n\dagger}_n(1)\rangle.
\ee
Requiring the normalization
$\langle O^{1_n{\bar 1}_n}_n(0)O^{1_n{\bar 1}_n\dagger}_n(1)\rangle=1$,
     we find the value of $\lambda_n$:
\be
\lambda_n=\left[{n(N-n)!\over N!}\right]^{-1/2}.
\ee
Let us now look at the three point function. For the case when the covering
surface is a sphere, the three point function is \cite{our}:
\bea\label{combinatorics}
&&\langle O^{1_n{\bar 1}_n}_n(0)O^{1_m{\bar 1}_m}_m(1)
O^{1_q{\bar 1}_q\dagger}_q(z)\rangle=
\frac{\sqrt{mnq(N-n)!(N-m)!(N-q)!}}{(N-s)!
\sqrt{N!}}\nonumber\\
&&\qquad\times\langle\sigma^{1_n{\bar 1}_n}_n(0)\sigma^{1_m{\bar 1}_m}_m(1)
\sigma^{1_q{\bar 1}_q\dagger}_q(z)\rangle.
\nonumber
\eea
with $s={1\over 2}(n+m+q-1)$.

Now we analyze the behavior of the combinatoric factors  for
arbitrary genus $g$ but in the limit where
          $N$ is taken to be large  while the orders of twist
operators ($m$, $n$
and $q$) as well as the  parameter $g$ are kept fixed.
There are $s$ different fields $X^i$ involved in the 3-point
function, and these fields can be selected in $\sim N^s$ ways.
Similarly the 2-point function of $\sigma_n$ will go as $N^n$ since
$n$ different fields are  to be selected. Thus the 3-point
function of normalized twist operators will behave as
\be
N^{s-{n+m+q\over 2}}= N^{-(g+{1\over 2})}
\label{efour}
\ee

Thus in the large $N$ limit the contributions from surfaces with high genus
will be suppressed, and  the leading order the answer can be obtained by
considering only contributions from the sphere ($g=0$). This is presicely the
case that we have analyzed in detail, and knowing the amplitude
for operators $\sigma$ one can easily extract the
leading order of the CFT correlation function:
\be
\langle O^{1_n{\bar 1}_n}_n(0)O^{1_m{\bar 1}_m}_m(1)
O^{1_q{\bar 1}_q}_q(z)\rangle=\sqrt{\frac{1}{N}}
{\sqrt{mnq}}\langle\sigma^{1_n{\bar 1}_n}_n(0)
\sigma^{1_m{\bar 1}_m}_m(1)\sigma^{1_q{\bar 1}_q}_q(z)
\rangle_{sphere}+O\left(\frac{1}{N^{3/2}}\right).
\ee

%
\section{Discussion}
\renewcommand{\theequation}{7.\arabic{equation}}
\setcounter{equation}{0}

In this paper we have computed the contribution to the 3-point
function for the case when the covering surface
$\Sigma$ is a sphere. The construction of the chiral operators and
the method of
computing correlators used only the fact that the CFT based on the
manifold $M$ had
$N=4$ supersymmetry; thus the computation is not restricted to
orbifolds that can be
obtained from free fields. The result we obtain is independent of the
details of $M$, and
thus exhibits a  `universal' property of CFTs arising from orbifolds $M^N/S^N$.

The sphere contribution  is the dominant contribution at large $N$,
but we can in principle calculate
the contribution of surfaces $\Sigma$ with $g>0$ to get an exact
result for finite $N$. Note that for a
given choice of orders of the chiral operators, the number of
different surfaces $\Sigma$ that can
contribute is finite, though it grows with the orders of the
operators. In the case of the bosonic
orbifold we had shown in \cite{our} that the  correlator for twist
operators on the $z$ plane for the
orbifold $M^N/S^N$ could be written in terms of the partition
functions $Z_g$ for the theory with one
copy of $M$ but on a worldsheet of genus $g$. For the supersymmetric
case we get the following
extension of this result: the correlator of the chiral operators
considered here can be written in
terms of the $Z_g$  and a finite number of derivatives of $Z_g$ with
respect to  moduli, where the
moduli arise from both the shape of the Riemann surface and from the
`current algebra moduli' that
couple to the $su(2)$ currents. This result follows from the fact that
we map the correlator of twist
operators to the covering surface $\Sigma$ as in the bosonic case,
but we get some
correlators of currents that
need to be computed on $\Sigma$. The current insertions can be
removed by using the
current algebra Ward identity,  but in the process we generate
derivatives with respect to the
moduli.

The supersymmetric orbifold that we have studied is expected to be a
point in the D1-D5 system
moduli space, and the latter system is dual to string theory on
$AdS_3\times S^3\times M$. In this
context we recall some observations that were made in \cite{our}
relating orbifold correlation
functions to the dual string theory. First, it was found that to get
a leading order 3-point function of
twist operators (which requires that the genus of $\Sigma$ be zero)
we must satisfy some
restrictions on the orders of the twists. These restrictions turn out
to be identical to the fusion rules
of the $su(2)$ WZW model, which restrict the 3-point functions of
string states at tree level in the
dual string theory. Secondly, we note that the orbifold theory has
expressed the correlators of twist
operators as a sum of contributions from different genus Riemann
surfaces, with the contribution of
higher genera surfaces being suppressed by $1/N^g$. This is
reminiscent of the genus expansion in the
dual string theory, where higher genus amplitudes are suppressed by a
similar factor ($1/\sqrt{N}$ is the
      coupling constant of the string modes).

It would be interesting to compare the 3-point functions that we have
computed to the 3-point
functions of supergravity field in the dual theory, to see if there
is any analogue of the
nonrenormalization that was found in the case of D-3 branes and the
dual theory on $AdS_5\times
S^5$. Some 3-point supergravity amplitudes have been computed in
\cite{ArutThei}, but these
correspond to scalar fields which are expected to be supersymmetry
descendents of the chiral fields
that we have worked with. Another supergravity 3-point correlation
function  was computed in
\cite{mihailescu}, and there is a significant agreement in overall
form between this result and the
correlators that we get. However the calculation of \cite{mihailescu}
used a step where a
symmetrization was performed over the three  fields involved in the
correlator, and  to reproduce
the result obtained from the orbifold computation we would have to
choose a somewhat different way
to symmetrize (instead of summing the squares of the three momenta we
would have to sum the
momenta and then square the result).  We hope to return to this
comparison at a later point.

The D1-D5 system for the black holes studied in
\cite{stromvafa}\cite{dasmathur}  arises
by wrapping the space direction of the D1-D5 CFT on a circle, with
periodic boundary
conditions for the fermions on this circle. Thus we need to study the
CFT in the Ramond
sector, rather than in the NS sector. We can compute correlation
functions in the
Ramond sector if we can compute correlators with insertions of spin
fields, since these
spin fields map the vacuum on the plane to the Ramond vacuum. (These
insertions of
spin fields are different from the spin fields that we have
encountered in the present
paper; the latter arose only after the map to the $t$ space, while
the former would be
inserted to change boundary conditions in the original $z$ space.) We
have computed
examples of correlators in the Ramond sector, which correspond to the
amplitude  for a
simple `black hole state' to absorb and re--emit a quantum; this
calculation will be
presented elsewhere \cite{lm}. In the dual theory on $AdS_3\times
S^3\times M$ the NS
and Ramond sector are connected by a spectral flow as well. It was suggested in
\cite{wadia} that a family of singular conical defect spacetimes
would represent this
spectral flow, but it was  argued recently in \cite{mathur} that the
generic spacetime in
the flow would in fact be smooth, and the spacetime deformation
corresponding to
spectral flow away from the NS vacuum was computed to first order .

\section*{Acknowledgments}

We are grateful to A. Jevicki, M. Mihailescu, S. Ramgoolan, and S.
Frolov for patiently explaining their results to us.   We also
benefited greatly from discussions with  S. Das, E.
D'Hoker,  C. Imbimbo,  D. Kastor, F. Larsen, E. Martinec,  S. Mukhi,
Z. Qiu,  L. Rastelli and  S.T. Yau.

\appendix
\section{$N=4$ superconformal algebra.}
\label{AppSuConf}
\renewcommand{\theequation}{A.\arabic{equation}}
\setcounter{equation}{0}
\bea
T(z)T(w)&=&\frac{\d T(w)}{z-w}+\frac{2T(w)}{(z-w)^2}+\frac{c}{2(z-w)^4},
\nonumber\\
J^i(z)J^j(w)&=&\frac{i\varepsilon^{ijk}J^k(w)}{z-w}+\frac{c}{12(z-w)^2},
\nonumber\\
T(z)J^i(w)&=&\frac{\d J^i(w)}{z-w}+\frac{J^i(w)}{(z-w)^2}, \nonumber\\
G^a(z){\tilde G}_b(w)&=&\frac{2T(w)\delta^a_b}{z-w}-
\frac{2{({\sigma}^i)^a}_b\d J^i(w)}{z-w}-
\frac{4{({\sigma}^i)^a}_bJ^i(w)}{(z-w)^2}+\frac{2c\delta^a_b}{3(z-w)^3},\\
T(z)G^a(w)&=&\frac{\d G^a(w)}{z-w}+\frac{3G^a(w)}{2(z-w)^2},\qquad
T(z){\tilde G}_a(w)=\frac{\d {\tilde G}_a(w)}{z-w}+
\frac{3{\tilde G}_a(w)}{2 (z-w)^2}, \nonumber\\
J^i(z)G^a(w)&=&-\frac{{(\sigma^i)^a}_b G^b(w)}{2(z-w)}, \qquad
J^i(z){\tilde G}_a(w)=\frac{{\tilde G}_{b}(w){(\sigma^i)^b}_a}{2(z-w)}\nonumber
\eea
The elements of $G^a$ and ${\tilde G}_{b}$ are related by complex conjugation:
\be
{\tilde G}_a(z)=\left(G^a(z)\right)^\dagger.
\ee
In terms of modes we have:
\bea
\left[L_m,L_n\right]&=&(m-n)L_{m+n}+\frac{c}{12}m(m^2-1)\delta_{m+n,0}
\nonumber\\
\left[J_m^i,J_n^j\right]&=&
i\varepsilon^{ijk}J^k_{m+n}+\frac{c}{12}m\delta_{m+n,0},\qquad
\left[L_m,J_n^i\right]=-nJ_{m+n}^i,\nonumber\\
\left\{G_r^a,{\tilde G}_{b,s}\right\}&=&2\delta^a_b L_{r+s}-
2(r-s){({\sigma}^i)^a}_b J^i_{r+s}+
\frac{c(4r^2-1)}{12}\delta^a_b\delta_{r+s,0},\\
\left\{G_r^a,G_s^b\right\}&=&0,\qquad
\left\{{\tilde G}_{a,r},{\tilde G}_{b,s}\right\}=0\nonumber\\
\left[L_m,G^a_r\right]&=&\left(\frac{m}{2}-r\right)G^a_{m+r},\qquad
\left[L_m,{\tilde G}_{a,r}\right]=\left(\frac{m}{2}-r\right){\tilde G}_{a,m+r}
\nonumber\\
\left[J_m^i,G_r^a\right]&=&-\frac{1}{2}{(\sigma^i)^a}_b G^b_{m+r},\qquad
\left[J_m^i,{\tilde G}_{a,r}\right]=\frac{1}{2}{\tilde
G}_{b,m+r}{(\sigma^i)^b}_a
\nonumber
\eea

As  was pointed out by
Schwimmer and Seiberg \cite{SchwimSeib}, there is a family of equivalent
representations of the superconformal algebra, which differ only by the
boundary conditions:
\be
J^{\pm}(z)=e^{\mp 2i\pi\eta}J^{\pm}(e^{2i\pi}z),\qquad
G^1(z)=-e^{i\pi\eta}G^1(e^{2i\pi}z),\qquad
{\tilde G}_2(z)=-e^{i\pi\eta}{\tilde G}_2(e^{2i\pi}z).
\ee
The equivalence can be established by the action of the spectral flow
\footnote{Since the R symmetry group for $N=4$ theory is $SU(2)$,
there is a three parametric family
of spectral flows. We will perform such flow only along $J^3$ direction}:
\bea
&&T_\eta (z)=T(z)-\frac{\eta}{z}J^3(z)+\frac{c\eta^2}{24z^2},\\
&&J^3_\eta(z)=J^3(z)-\frac{c\eta}{12z},\qquad
J^\pm_\eta(z)=z^{\mp\eta}J^\pm(z),\\
&&G^1_\eta(z)=z^{\frac{\eta}{2}}G^1(z),\qquad
G^2_\eta(z)=z^{-\frac{\eta}{2}}G^2(z),\\
&&{\tilde G}_{1\eta}(z)=z^{-\frac{\eta}{2}}{\tilde G}_1(e^{2i\pi}z),\qquad
{\tilde G}_{2\eta}(z)=z^{\frac{\eta}{2}}{\tilde G}_2(e^{2i\pi}z)
\eea
In particular, if we started with ``vacuum state'' $|\chi\rangle$:
\be
\oint \frac{dz}{2\pi i}z^{p+1}T(z)|\chi\rangle=
\oint \frac{dz}{2\pi i}z^{p}J^3(z)|\chi\rangle=0,\qquad p\ge 0.
\ee
in $\eta=0$ sector, then the charges in $\eta$ sector are given by:
\be
\oint \frac{dz}{2\pi i}zT_\eta(z)|\chi\rangle=\frac{c\eta^2}{24},\qquad
\oint \frac{dz}{2\pi i}J^3_\eta(z)|\chi\rangle=-\frac{c\eta}{12},\qquad c=6.
\ee
To have a well defined expression for $J(z)$
one has to choose an odd integer $\eta$, but for odd $\eta$ the fermionic
operators become integer--moded, i.e. we obtain a Ramond sector of the theory.
We will pick the simplest possible
value: $\eta=1$.
Thus, starting point of our flow (i.e. NS sector) corresponds to $\eta=0$,
while at
$\eta=1$ we get a Ramond sector. We will also need relations between modes in
NS and R sectors (primed modes are from the R sector):
\bea
&&T'_n=T_n-J^3_n+\frac{c}{24}\delta_{n,0},\qquad
(J^3)'_n=J^3_n-\frac{c}{12}\delta_{n,0},\qquad
(J^\pm)'_n=J^\pm_{n\mp 1},\\
&&(G^1)'_n=G^1_{n+\frac{1}{2}},\qquad
(G^2)'_n=G^2_{n-\frac{1}{2}},\qquad
{\tilde G}'_{1,n}={\tilde G}_{1,n-\frac{1}{2}},\qquad
{\tilde G}'_{2,n}={\tilde G}'_{2,n+\frac{1}{2}}\nonumber
\eea
In the NS sector there was a bound on a maximal possible charge: $|j|\le h$,
which saturated only for (anti)chiral fields. In particular, chiral fields had
$h=j$. In the Ramond sector this bound becomes: $h'\ge \frac{c}{24}$ and it
is saturated only by images of chiral fields.

\section{Three point function for the bosonic orbifold.}
\label{AppBoson}
\renewcommand{\theequation}{B.\arabic{equation}}
\setcounter{equation}{0}
In this appendix we will summarize the results of \cite{our}, where the
three
point function on the bosonic $S^N$ orbifold was calculated. In
particular, in
\cite{our} we evaluated the contribution to the three point function for
the
case where the covering space is a sphere. In this case the map
corresponding
to the three point function
\be\label{App3ptEqn}
\langle\sigma_n(0)\sigma_m(a)\sigma_q(\infty)\rangle
\ee
is given by:
\be
\label{mapjac}
z=at^nP_{d_1-n}^{(n,-d_1-d_2+n-1)}(1-2t)
\left[P_{d_2}^{(-n,-d_1-d_2+n-1)}(1-2t)\right]^{-1}.
\ee
This map involves Jacobi polynomials $P_{n}^{(\alpha,\beta)}(x)$ and the
values of $d_1$ and $d_2$ are defined as
\be
d_1=\frac{1}{2}(n+m+q-1),\qquad d_2=\frac{1}{2}(n+m-q-1).
\ee
One can easily show that the asymptotic behavior of this map is given by
(\ref{asym1}), (\ref{asym2}), (\ref{asym3}). We recall that the three
point
function (\ref{App3ptEqn}) was evaluated in \cite{our} by considering the
conformal anomaly for the map (\ref{mapjac}). More precisely, this three point
function
can be written in terms of the Liouville action describing the map
(\ref{mapjac})
and the Liouville actions corresponding to two point functions
$\langle\sigma_n(0)\sigma_n(z)\rangle$:
\bea\label{AppDefCAnom}
&&\frac{\langle\sigma_n(0)\sigma_m(a)\sigma_q(\infty)
\rangle}{\langle\sigma_q(0)\sigma_q(\infty)\rangle}=
\exp\left(S_L(\sigma_n(0),\sigma_m(a),\sigma_q(\infty))-
S_L(\sigma_q(0)\sigma_q(\infty))\right)\nonumber\\
&&\qquad\times\exp\left(-\frac{1}{2}S_L(\sigma_n(0),\sigma_n(1))-
{1\over 2}S_L(\sigma_m(0),\sigma_m(1))+{1\over 
2}S_L(\sigma_q(0),\sigma_q(1))\right)
\nonumber\\
&&\qquad\equiv|C_{n,m,q}|^{2c}|a|^{-\frac{c}{12}(n+m-q-1/n-1/m+1/q)}
\eea
  $C_{n,m,q}$  is the fusion coefficient of $\sigma_n$ and $\sigma_m$ 
to $\sigma_q$.

The evaluation of the three
point function leads to the following result \cite{our}:
\bea\label{CformAnom}
\log| C_{n,m,q}|^2&=&\frac{1}{6}\log\left(\frac{q}{mn}\right)
-\frac{n-1}{12}\log n-\frac{m-1}{12}\log m +\frac{q-1}{12}\log(q)
\nonumber\\
&-&\frac{n-1}{12n}\log\left(\frac{d_1!d_2!}{n!(n-1)!}\
\frac{(d_1-m)!}{(d_1-n)!}\right)\nonumber\\
&-&\frac{m-1}{12m}\log\left(\frac{d_1!d_2!}{m!(m-1)!}\
\frac{(d_1-n)!}{(d_1-m)!}\right)\nonumber\\
&+&\frac{q-1}{12q}\log\left(\frac{(q-1)!d_2!}{(d_1-n)!(d_1-m)!}\
\frac{(d_1-d_2)!)}{d_1!}\right)\\
&+&\frac{1}{3}d_2(d_2-1)\log 2-\frac{d_2}{6}\log n+\frac{1}{3}\log{\cal
D}+
\frac{3d_2-4}{6}\log d_2!\nonumber\\
&-&\frac{d_2}{6}\log\left[\frac{d_1!}{n!(d_1-n)!}\right]-
\frac{n+d_2-1}{6}\log\frac{(n-1)!}{(n-d_2-1)!}\nonumber\\
&-&\frac{d_1-d_2+3}{6}\log\frac{(d_1-d_2)!}{d_1!}-
\frac{d_1+d_2-n}{6}\log\frac{(d_1+d_2-n)!}{(d_1-n)!}.\nonumber
\eea
This expression involves the discriminant of the Jacobi polynomials ${\cal
D}$,
        which is known as a following finite product \cite{szego}:
\bea\label{discriminant}
{\cal D}&\equiv&D_{d_2}^{(-n,-d_1-d_2+n-1)}=2^{-d_2(d_2-1)}\\
&\times&\prod_{j=1}^{d_2} j^{j+2-2d_2}
(j-n)^{j-1}(j-d_1-d_2+n-1)^{j-1}(j-d_1-1)^{d_2-j}.\nonumber
\eea

Note that the combination of Liouville actions entering
(\ref{AppDefCAnom})
is the combination needed for SUSY orbifold (\ref{overall}). Thus we can
take
the result for $|C_{n,m,q}|$ as contribution of the conformal anomaly even
for
supersymmetric case.

\section{Calculation of ${\hat C}^{+,-,+}_{n,m,q}$.}
\label{AppendC}
\renewcommand{\theequation}{C.\arabic{equation}}
\setcounter{equation}{0}

In section \ref{Sect3Pt} we have calculated ${\hat C}^{-,-,-}_{n,m,q}$ and
presented
the result for general reduced fusion coefficient
${\hat C}^{1_n,1_m,1_q}_{n,m,q}$.
In this section we will assume the relation (\ref{PostuLate}) and, using it,
we will evaluate another fusion coefficient ${\hat
C}^{+,-,+}_{n,m,q}$ which will give
another case of the result (\ref{TheFinal}).
Using the general expression (\ref{theResult}), one can find a ratio:
\be
\frac{C^{+(d_2),-,+}_{n,m,q}}{C^{-(d_2),-,-}_{n,m,q}}=
\frac{n!(n-1)!q!(q-1)!}{\left(d_1!(d_1-m)!\right)^2}
\left(\frac{(n-d_2)(n-d_2+1)}{n(n+1)}\right)^{1/2}
\frac{L^{d_2}_{n+1,m-1}}{L^{d_2}_{n-1,m-1}}.
\ee
To evaluate a ratio of limits one can use the fact that for an integer $l>1$:
\be
\prod_{j=1}^d\left[\frac{j-l-2}{j-l}\right]^{1-j}=
\left(\frac{(l-d-1)!}{(l-1)!}\right)^2\frac{(l-d)^{d+1}(l-d+1)^d}{l}.
\ee
Then one gets:
\be
\frac{L^{d_2}_{n+1,m-1}}{L^{d_2}_{n-1,m-1}}=
\left[\frac{(n-d_2-1)!(m+n-d_2-1)!}{(m+n-2d_2-1)!(n-1)!}\right]^2
\frac{n-d}{n}\frac{m+n-2d_2-1}{m+n-d_2-1}
\ee
To define the reduced fusion coefficient we also need the relation between
appropriate $3j$ symbols:
\be
\left(
\begin{array}{ccc}
\frac{n+1}{2}&\frac{m-1}{2}&\frac{q+1}{2}\\
\frac{q-m+2}{2}&\frac{m-1}{2}&-\frac{1+q}{2}
\end{array}
\right)=\left[\frac{q(q+1)}{(d_1+1)(d_1+2)}\right]^{1/2}
\left(
\begin{array}{ccc}
\frac{n-1}{2}&\frac{m-1}{2}&\frac{q-1}{2}\\
\frac{q-m}{2}&\frac{m-1}{2}&\frac{1-q}{2}
\end{array}
\right),
\ee
which can be deduced from the general expression:
\be
\left(
\begin{array}{ccc}
a&b&c\\
c-b&b&-c
\end{array}
\right)=\left(\frac{[2b]![2c]!}{[a+b+c+1]![b+c-a]!}\right)^{1/2}.
\ee
Collecting all this information together, one finds the reduced fusion
coefficient:
\bea
{\hat C}^{+,-,+}_{n,m,q}&=&{\hat C}^{-,-,-}_{n,m,q}
\left[\frac{(d_1+1)(d_1+2)}{(q+1)(n+1)}\frac{(d_1-m+1)(d_1-m+2)}{nq}
\right]^{1/2}
\frac{d_1-m+1}{d_1}\nonumber\\
&=&\left(\frac{\left(d_1-m+1\right)^2}{mnq}
\frac{(d_1+2)!\ (d_1-q)!(d_1-n)!(d_1-m+2)!}
{(n+1)!~(m-1)!~(q+1)!}\right)^{1/2},
\eea
which agrees with (\ref{TheFinal}).

Proceeding in a similar manner for other values of the
indices $1_n,1_m,1_q$ we arrive at (\ref{TheFinal}).

\end{document}